\documentclass[preprint,12pt]{elsarticle}

\usepackage{amssymb}
\usepackage{amsthm}
\usepackage{amsmath}

\usepackage{array}

\usepackage[section]{placeins} % force figures etc to appear in section

\usepackage{xcolor}

\usepackage{algorithm}%
\usepackage{algorithmicx}%
\usepackage{algpseudocode}%

\usepackage{longtable}

\usepackage{lineno}

\newcommand{\PPhi}{\overline{\Phi}}

\newcommand{\bX}{\mathbf{X}}

\newcommand{\bR}{\mathbf{R}}

\newcommand{\bZ}{\mathbf{Z}}

\newcommand{\bK}{\mathbf{K}}
\newcommand{\bL}{\mathbf{L}}

\newcommand{\bu}{\mathbf{u}}

\newcommand{\x}{\boldsymbol{x}}

\newcommand{\bphi}{\boldsymbol{\phi}}

\newcommand{\real}{\mathbb{R}}

\newcommand{\hp}{\alpha}
\newcommand{\dhp}{\Delta\alpha}
\newcommand{\hpb}{\boldsymbol{\alpha}}

\newcommand{\df}{G}

\newcommand{\phibar}{\bar{\phi}}

\newcommand{\ones}{\mathbf{1}}

\usepackage{titlesec}
\titleformat*{\subsection}{\bfseries}

\journal{Structural Safety}

\graphicspath{{../matlab/png}}

\begin{document}
%\linenumbers

\begin{frontmatter}

\title{Reliability Sensitivity with Response Gradient}

%\author{Siu-Kui Au}

\author[1]{Siu-Kui Au}
\author[2]{Zi-Jun Cao}

\affiliation[1]{organization={School of Civil and Environmental Engineering, Nanyang Technological University}, 
%addressline={50 Nanyang Avenue}, 
%            city={Singapore},
%            postcode={639798}, 
%            state={},
            country={Singapore}
            }

\affiliation[2]{organization={MOE Key Laboratory of High-Speed Railway Engineering, Institute of Smart City and Intelligent Transportation, Southwest Jiaotong University}, 
%addressline={}, 
            city={Chengdu},
%            postcode={}, 
%            state={},
            country={China}
            }

%==============================================
% abstract
%==============================================
\begin{abstract}
Engineering risk is concerned with the likelihood of failure and the scenarios when it occurs. 
The sensitivity of failure probability to change in system parameters is relevant to risk-informed decision making. 
Computing sensitivity is at least one level more difficult than the probability itself, which is already challenged by a large number of input random variables, rare events and implicit nonlinear `black-box' response. 
%
%It involves resolving the differentation and integration with respect to the sensitivity parameters and input random variables, respectively.   
%
Finite difference with Monte Carlo probability estimates is spurious, requiring the number of samples to grow with the reciprocal of step size to suppress estimation variance. 
Many existing works gain efficiency by exploiting a specific class of input variables, sensitivity parameters, or response in its exact or surrogate form.  
For general systems, this work presents a theory and associated Monte Carlo strategy for computing sensitivity using response values and gradients with respect to sensitivity parameters.  
It is shown that the sensitivity at a given response threshold can be expressed via the expectation of response gradient conditional on the threshold.  
Determining the expectation requires conditioning on the threshold that is a zero-probability event, but it can be resolved by the concept of kernel smoothing. 
The proposed method offers sensitivity estimates for all response thresholds generated in a single Monte Carlo run. 
It is investigated in a number of examples featuring sensitivity parameters of different nature.
As response gradient becomes increasingly available, it is hoped that this work can provide the basis for embedding sensitivity calculations with reliability in the same Monte Carlo run.  
\end{abstract}

%==============================================
% highlights
%==============================================
\begin{highlights}

\item Reliability sensitivity as expectation of response gradient

\item Kernel smoothing resolves conditioning on zero-probability event

\item Sensitivity embedded in Monte Carlo run 

\item Applicable regardless of the type of sensitivity parameter

\end{highlights}

%==============================================
% keywords
%==============================================

\end{frontmatter}

%% main text
%===============================================
\section{Introduction}
%===============================================

%--------------------------------------------------
% introduce general terms and referneces
%--------------------------------------------------
Engineering risk has many facets that cover the likelihood of failure, the scenarios when failure occurs, life-cycle costs, etc. 
Modern systems are increasingly complex, demanding algorithms that are scalable with the number of input random variables and nonlinearity in input/output relationship, and remain feasible for rare events with low probability. 
See monographs on engineering reliability~\cite{Ditlevsen1996,melchers2017,DerKiureghian2022}, a recent a review on structural reliability~\cite{Ellingwood2025}, and a journal special issue on methodology for complex systems~\cite{Teixeira2023}.
%
% spare ref: Klppelberg2014
%%
%To cope with complex system response that can be practically a `black-box', computational methods are increasingly `Monte Carlo' in nature, i.e., estimation by averaging over pseudo-randomly generated samples of input random variables.
%%
%See classical references on common strategies~\cite{robert2004} and random sample generation from standard distributions~\cite{devroye1986}. 
%%
Beyond a given scenario, assessing the change of risk measures or expected performance to perturbations in system parameters/assumptions is relevant to risk-informed decision making, e.g., establishing confidence in conclusions, hypothesis testing of assumptions and reliability-based design; see~\cite{Hu2024} a recent review.   
The concept is also practiced in non-engineering fields such as finance, e.g., `stress testing' of credit risk and the `Greeks' (sensitivity) of stock option price; see~\cite{Kosowski2015} a textbook.

%----------------------------
% problem setup
%----------------------------
This work is concerned with computing reliability sensitivity for systems when the response and its gradient with respect to (w.r.t.\@) `sensitivity parameters' are available, but otherwise they can remain as a `black-box'. 
Let
\begin{align}\label{eq:f_def}
Y = f(\bX,\hp)
\end{align}
be a scalar response of interest that depends on the random input vector $\bX\in \real^n$ and a deterministic (scalar) sensitivity parameter $\hp$ through the response function $f$.
Failure is defined as the exceedance of $Y$ over a given threshold $y$, with probability
\begin{align}\label{eq:F_def}
F(y,\hp) = P(Y\ge y) = \int_{f(\x,\hp)\ge y} q(\x)\, d\x
\end{align}
where $q$ denotes the probability density function (PDF) of $\bX$.
Although not explicitly indicated, $P(Y\ge y)$ depends implicitly on $\hp$ through \eqref{eq:f_def}.
Viewed as a function, $F(\cdot,\hp)$ is the `complementary cumulative distribution function' (CCDF) of $Y$ for a given $\hp$.
Reliability sensitivity is generally concerned with how $F$ changes with $\hp$.
In this work we focus on the derivative $\partial F/\partial \hp$.

%---------------------------------------------
\subsection{Notes on problem context}
%---------------------------------------------
As is conventional, despite the name `reliability', $F$ instead of $1-F$ is considered because the latter is close to 1 in applications, masking significant digits in quantitative analysis or interpretation. 
Considering a scalar $Y$ leads to no loss of generality because the critical response of a system comprising sub-systems connected in series or parallel, or any combination, can be defined as the max or min of component responses cascaded in the same manner.  
Having said that, there are strategies that allow one to generate reliability estimates for multiple responses with less computations than repeated runs~\cite{Li2015,Xia2025}, but they are out of the present scope. 
As indicated by the notation, it is assumed that the PDF $q$ does not depend on $\hp$. This is merely to standardize the mathematical context. It does not lead to any loss of generality because parameters that affect the PDF can be absorbed into the definition of $f$.
Having said that, there are works that explore efficiency gain by limiting $\hp$ to `distribution parameters', i.e., those affecting $q$ only; see literature review later in Section~\ref{sec:review}.
The assumption of a scalar $\hp$ is just to simplify notation and leads to no loss of generality either, because the gradient of $F$ w.r.t.\@ multiple parameters is simply a vector of the individual derivatives as long as the parameters are `free', i.e., they do not depend on each other and are not subjected to any equality constraint. 
%

%-----------------------------------------------------------------------------------
\subsection{Fundamental challenge and state of the art}\label{sec:review}
%-----------------------------------------------------------------------------------
From first principles the sensitivity $\partial F/\partial \hp$ involves the derivative of $F$ w.r.t.\@ $\hp$, while $F$ in turn involves a multivariate integral over the failure domain $\{\x\in \real^n: f(\x,\hp)>y\}$ that is only implicitly known when $f(\cdot,\hp)$ is a black-box. 
Efficient calculation of $F$ for a given $\hp$ belongs to a reliabilty estimation problem. It involves resolving the integral, e.g., locating the failure domain and accounting for contributions there.
The key challenges are 1) high dimension $n$ that renders deterministic search algorithms unsustainable and intuitions in low dimensions inapplicable or misleading; 2) small probability ($F$) that makes direct Monte Carlo prohibitive and poses a wide knowlege gap for machine-learning to set off from prior knowledge; 3) implicit nonlinear $f(\cdot,\hp)$ that makes approximation (e.g., Taylor series) or machine-learning difficult, not to mention the cost of each function evaluation. 
These challenges are well-awared in the reliability literature, and research is still growing. 
State-of-the-art strategies for general applications are mostly Monte Carlo in nature, i.e., estimation by averaging over pseudo-randomly generated samples of input random variables.
%
%See classical references on common strategies~\cite{robert2004} and random sample generation from standard distributions~\cite{devroye1986}. 
% 
% spare ref: 
%  Cheng2025 on directional importance sampling
%  Patelli2016 on line samping
%
There is a large variety of methods, e.g., direct Monte Carlo that is most robust but prohibitive for rare events; importance sampling when a proposal PDF can be constructed with heuristics or domain-knowledge in various forms~\cite{Papaioannou2016,Deo2023}; Subset Simulation (SS)~\cite{Au2014,Zuev2021} (see also the end of Section~\ref{sec:kernel_ss}) that is less demanding on prior knoweldge, though at the expense of a machine-learning curve from frequent to rare events that can get tougher along the way; Line Sampling~\cite{deAngelis2015,Dang2024_LS} that exploits the probability along different lines of directions in the input variable space; and recent improvements with data science tools such as `Krigging'~\cite{Xin2025,Dai2025} that originates from geostatistics, active learning~\cite{Dang2024,Rajak2024} that improves the selection of sample points, and neutral network~\cite{Xu2023,Shi2025} that offers flexiblity for surrogates and sample generation beyond developers' mathematical facility.

% fundamental difficulty with sensitivity
%----------------------------------------------------------------
\subsubsection{Finite difference}\label{sec:fd}
%----------------------------------------------------------------
Computing $\partial F/\partial \hp$ is at least one level more difficult than $F$, as it needs to resolve both the derivative (w.r.t.\@ $\hp$) and integral (over $\x$). 
Requesting the sensitivity for multiple values of $y$, as is possible for $F$ in a single Monte Carlo run, adds to one's plate. 
%
%Integration by parts or Green's theorem (for $n\ge 2$) does not help, as the derivative and integral are over different spaces. 
%----------------------------------------------------------------
% central difference of Monte Carlo
%----------------------------------------------------------------
One last resort is finite difference, e.g., $\partial F/\partial \hp \approx  [\widetilde{F}(\hp+\Delta \hp) - \widetilde{F}(\hp-\Delta)]/2\Delta \hp$, where $\widetilde{F}(\hp\pm \Delta \hp)$ denotes the Monte Carlo estimates of $F$ at $\hp\pm \Delta \hp$ but based on the same set of $N_s$ (say) samples of $\bX$, i.e., `common random numbers' (CRN)~\cite{Glasserman1992}.
Steming from central difference, the estimate has a bias of $O(\Delta\hp^2)$. 
In addition to the bias, for typical situations where $\widetilde{F}(\hp +\Delta \hp)$ and $\widetilde{F}(\hp -\Delta \hp)$ have a correlation of $O(\Delta \hp)$, the variance of the estimate is $O(1/N_s\Delta \hp)$, requiring at least $N_s = O(\Delta \hp^{-1})$ samples to suppress. This is a fundamental issue with finite difference of Monte Carlo (even with CRN), and is the original of many suprisingly spurious results when the issue is not awared of. 
%

%----------------------------------------
\subsubsection{Distribution parameters}\label{sec:distpar}
%----------------------------------------
In view of the issue with finite difference, `stochastic gradient estimation' is an area that explores functions whose expectation is equal or at least tends to the target gradient, in our case the reliability sensitivity. See~\cite{Fu1997} a monograph in the area of discrete event simulation.
Inevitably, efficiency is gained at the expense of limited scope or approximation.
%
%----------------------------------------------------------------
% works that assume alpha = distribution parameter
%----------------------------------------------------------------
Some existing works assume that $\hp$ affects $q$ but not $f$ so that the differentiation falls on $q$ only, i.e., $\partial F/\partial \hp = \int_{f(\x,\hp)\ge y} \partial q/\partial \hp\, d\x$. In this case strategies have been developed to evaluate $\partial F/\partial \hp$ as an integral or expectation, e.g., $\partial F/\partial \hp = E\left[I\left(f(\bX,\hp)\ge y\right) S(\bX,\hp)\right]$ where $\bX\sim q$; $S = q^{-1}\partial q/\partial \hp$ is a `score function'; and $I(\cdot)$ is the indicator function, equal to 1 when the argument is true and zero otherwise. 
Issues with evaluating $F$ carry over to the expectation $E[\cdot]$. 
%--------------
% Zijun's 
%--------------
%With this, the reliability sensitivity with respect to $\hp$ that only affects $q$ can be evaluated from a domain integral by the definition of expectation. Similar to reliability analysis for calculating $F$, the expectation herein can be calculated from direct Monte Carlo method, which can be prohibitive for sophisticated numerical models. 
% spare ref: 
%  \cite{Wu1993} on adaptive importance sampling
%
Variance reduction methods have been developed using, e.g., 
line sampling~\cite{Lu2008,Valdebenito2018,Valdebenito2019}, Subset Simulation~\cite{Song2009,Jensen2015}, and moving particles method~\cite{Proppe2021}. 
%
%These studies considered distribution parameters of applications. Valdebenito et al. (2019) explored the possibility of evaluating the reliability sensitivity with respect to parameters (e.g., correlation length) of random field models characterizing soil spatial variability with applications to pile design and seepage analysis based on stochastic finite element models and line sampling. 
%

%----------------------------------------
\subsubsection{General parameters}\label{sec:general_par}
%----------------------------------------
Expressing $\partial F/\partial \hp$ as an expectation that is conducive to Monte Carlo is more challenging in the general case where $\hp$ can affect the response function as well. 
Differentiating $F$ in \eqref{eq:F_def} under integral sign, e.g., using Leibniz rule, is not trivial, because the integral is over the failure domain that depends implicitly on $(\x,\hp)$. 
Writing $F = \int I\left(f(\x,\hp)\ge y\right) q(\x) d\x$ removes this dependence, but still does not allow differentiation using ordinary rules, however, because changing $\hp$ moves the failure (boundary) surface $\{\x\in \real^n: f(\x,\hp) = y\}$ that has a non-trivial effect on $F$. In fact, $\partial I(\cdot)/\partial \hp$ is zero except at the failure surface where it does not even exist. 
%
%The same issue applies when one explores Monte Caro with $F = E\left[I\left(f(\bX,\hp)\ge y\right)\right]$ where $\bX\sim q$. 
%
In this regard, it is possible to write in terms of a surface integral, as shown in Theorem~21 of~\cite{Breitung1994}. Applying the theorem with a limit state function $g(\x,\hp) = y-f(\x,\hp)$ and considering $q$ a function of $(\x,\hp)$, one obtains 
\begin{align}\label{eq:dF_surface}
\frac{\partial F}{\partial \hp} = 
\int_{f(\x,\hp)\ge y} \frac{\partial q}{\partial \hp} d\x
+ \int_{f(\x,\hp)=y} \frac{\partial f/\partial \hp}{\|\partial f/\partial \x\|} q(\x) dS(\x)
\end{align}
where the second integral is over the failure surface with differential element $dS(\x)$; $\partial q/\partial \hp$, $\partial f/\partial \hp$ and $\partial f/\partial \x$ are all functions of $(\x,\hp)$; $\|\cdot\|$ denotes the Euclidean norm. 
The first and second term in \eqref{eq:dF_surface} account for the influence of $\hp$ on $q$ and $f(\x,\hp)$, respectively. 
The first term is what was treated in methods that limit $\hp$ to distribution parameters. For its surface integral nature, the second term is more difficult to handle. 
Monte Carlo is not directly applicable because of the need to sample $\bX$ on the failure surface. 
In view of this, a `weak approach' was developed, where an augmented CDF was introduced as a smooth approximation to the indicator function to avoid the surface integral and allow Monte Carlo averaging.
See~\cite{Papaioannou2018} that employs sequential importance sampling to adaptively determine the augmented CDF.  
See~\cite{Torii2020} and~\cite{Torii2021} for discussion of score function method, weak approach, and CRN. 
See also~\cite{Chiron2023} a review on reliability sensitivity.

\section{Key contributions and organization of this work}
%===============================================
Advancing the state of the art, this work develops a theory and associated Monte Carlo strategy for reliability sensitivity based on response function values and derivatives w.r.t.\@ sensitivity parameters.  
To facilitate understanding, we summarize in this section the key contributions and outline the organization of this work.

%------------------------------------------------------------------
\subsection{Theory}\label{sec:key_theory}
%------------------------------------------------------------------
As the key theoretical contribution, we first derive in Section~\ref{sec:theory} the following formula for sensitivty: 
\begin{align}\label{eq:dF_intg}
\frac{\partial F}{\partial \hp} 
= \int t\, p(\df=t,Y=y)\, dt
\end{align}
where
\begin{align}\label{eq:G}
\df = \frac{\partial f(\bX,\hp)}{\partial \hp} 
\end{align}
is the (random) response gradient, and $p(\df = t,Y=y)$ denotes the joint PDF of $\{\df,Y\}$ at $\{t,y\}$. 
By writing $p(\df=t,Y=y) = p(\df=t|Y=t)\,p(Y=t)$, \eqref{eq:dF_intg} can be written as
\begin{align}\label{eq:dF_E}
\frac{\partial F}{\partial \hp} 
= p(Y=y) \, 
E[\df |Y = y]
\end{align}
which expresses the sensitivity in an intuitve form in terms of the PDF of $Y$ and the conditional expectation of $\df$ given that $Y=y$:
\begin{align}
E[\df|Y=y] = \int t\, p(G=t|Y=y)\, dt
\end{align}
%------------------------------------
% significance of theory
%------------------------------------
Equation \eqref{eq:dF_intg} is a signficant theoretical result, as it resolves the implicit dependence of $P(Y\ge y)$ on $\hp$, now expressing explicitly in terms of an integral that can be interpreted as an expectation.
From first glance it may appear that the result is merely an application of chain rule of differetiation, i.e., $\partial F/\partial \hp = \partial F/\partial y \cdot \partial y/\partial \hp$, where $\partial F/\partial y$ gives the PDF $p(Y=y)$ (which is correct) and $\partial y/\partial \hp$ `somehow' becomes the conditional expectation $E[G|Y=t]$. This looks intuitive but is not correct. In the first place $y$ is a fixed threshold, and hence not a function of $\hp$. 
%
%As will be seen in Section~\ref{sec:theory}, the proof of \eqref{eq:dF_intg} involves evaluating derivative from first principles, i.e., as the limit of the ratio of perturbations. 

%
Considering the general context of the sensitivity problem in this work, \eqref{eq:dF_intg} and its intuitive form in \eqref{eq:dF_E} are remarkably simple. 
Effectively, the derivative of probability, which is not trival and is the key challenge of a sensitivity problem, is now passed down to the response function which can be calculated in a deterministic manner. As it turns out, the conditional expectation correctly reflects the probability nature of the subject under differentiation. 
The proof is not trivial, however, as it involves evaluating the derivative of $P(Y\ge b)$ from first principle of Calculus, i.e., as the limit of the ratio of perturbations. 
See also Section~\ref{sec:related} that compares~\eqref{eq:dF_intg} with \eqref{eq:dF_surface}.

%------------------------------------------------------------------
\subsection{Monte Carlo with kernel smoothing}\label{sec:key_comp}
%------------------------------------------------------------------
Beyond the rare event challenge that is typical in the estimation of $F$, the additional challenge associated with $\partial F/\partial \hp$ lies in the integral in \eqref{eq:dF_intg} when the joint PDF of $\{\df,Y\}$ is not known, or the conditional expectation $E[\df|Y=y]$ in \eqref{eq:dF_E}. 
For the latter, Monte Carlo averaging requires samples conditional on the zero-probability event $\{Y=y\}$, which is not feasible. 
As the second key contribution, we have developed a simple approach that allows $\partial F(y,\hp)/\partial \hp$ to be computed together with the CCDF $F(y,\alpha)$ {\it for all sample values of $y$} generated in a single Subset Simulation (SS) run. 
The difficulty with zero probability event is overcome using the concept of `kernel smoothing', where Monte Carlo samples in the neighborhood of $\{Y=y\}$ are exploited for estimation at the expense of a bias that is nevertheless controllable. 
%----------------------------
% direct Monte Carlo
%----------------------------
Based on \eqref{eq:dF_intg}, we show in Section~\ref{sec:kernel_mc} that the sensitivity can be estimated using direct Monte Carlo samples of $Y_{k} = f(\bX_k,\hp)$ and $\df_k = \partial f(\bX_k,\hp)/\partial \hp$ with $\bX_k\sim q$: 
\begin{align}\label{eq:dF_mc}
\frac{\partial F(y,\hp)}{\partial \hp} \approx 
\sum_{k=1}^{N} \df_{k}\, w^{-1} K\left(\frac{Y_{k}-y}{w}\right)
\end{align}
where $K(\cdot)$ is a `kernel function', e.g., standard Normal PDF, and  $w$ is the `kernel width' (which can depend on $y$) that is chosen to balance over estimation bias (the narrower the better) and variance (the wider the better).

%----------------------------
% SS
%----------------------------
For the same difficulty with risk analysis for rare events, \eqref{eq:dF_mc} has large variance for small $F$ where Monte Carlo samples are lacking.   
Using SS for generating rare event samples, we show in Section~\ref{sec:kernel} that the sensitivity can be estimated by 
\begin{align}\label{eq:dF_SS}
\frac{\partial F(y,\hp)}{\partial \hp} \approx 
\sum_{i=0}^{m-1} P_i N_i^{-1} \sum_{k}^{N_i} \df_{ik}\, w_{i}^{-1} K\left(\frac{Y_{ik}-y}{w_{i}}\right)
\end{align}
%
%\begin{align}\label{eq:dF_SS}
%\frac{\partial F(y,\hp)}{\partial \hp} \approx 
%\sum_{i=0}^{m} P_i N_i^{-1} \sum_{k}^{N_i} \df_{ik}\, w_{i}^{-1} K\left(\frac{Y_{ik}-y}{w_{i}}\right)
%\end{align}
%%
%
%The $O(w^2)$ indicates that the sensitivity can be estimated by the sum with a second order bias, where $w$ denotes a representative scale among the $w_{ik}$'s. 
%
In \eqref{eq:dF_SS}, the subscript $ik$ denotes the $k$th sample in the `threshold bin' $B_i$ in SS. 
See Table~\ref{tab:bins} for the definition of $B_i$, its probability $P_i$ (on a sample basis), and the number of samples $N_i$ in the bin.  
They are defined in the same way as in \cite{Au2007_augment} except that $B_{m-1}$ here combines $B_{m-1}$ and $B_m$ there; see (3)-(5) in Section~2 there.
The kernel width $w_i$ can depend on the bin $i$, or even $y$ (not indicated). 
See \eqref{eq:scott} one simple choice based on Normal kernel.

%=====================
% Table: Bins in SS
%=====================
\begin{table}[h]
\caption{Definition of threshold bins $\{B_i\}_{i=0}^{m-1}$ in Subset Simulation}
\label{tab:bins}
\begin{center}
\begin{tabular}{m{10em} m{10em} m{10em}}
%---------------------------------------------------------
\hline
Bin & Probability & No. of samples\\
\hline
%---------------------------------------------------------
%\vbox{\begin{flushleft}
%$B_i = \{y_i \le Y < y_{i+1}\}$\\
%$0\le i\le m-2$ 
%%
%\end{flushleft}}
$B_i = \{y_i \le Y < y_{i+1}\}$
& $P_i = p_0^i (1-p_0)$ & $N_i = (1-p_0) N$\\
$0\le i\le m-2$\\
\hfill\\
%\hline
%---------------------------------------------------------
$B_{m-1} = \{Y\ge  y_{m-1}\}$ & $P_{m-1} = p_0^{m-1}$ & $N_{m-1} = N$\\
\hline
%---------------------------------------------------------
$y_0 = -\infty$
\end{tabular}
\end{center}
\end{table}

%##################
\FloatBarrier
%##################

%%=====================
%% Table: Bins in SS
%%=====================
%\begin{table}[h]
%\caption{Definition of threshold bins $\{B_i\}_{i=0}^m$ in Subset Simulation}
%\label{tab:bins}
%\begin{center}
%\begin{tabular}{m{10em} m{10em} m{10em}}
%%---------------------------------------------------------
%\hline
%Bin & Probability & No. of samples\\
%\hline
%%---------------------------------------------------------
%$B_0 = \{Y\le y_1\}$ & $P_0 = 1-p_0$ & $N_1 = (1-p_0) N$\\
%%\hline
%%---------------------------------------------------------
%\vbox{\begin{flushleft}
%$B_i = \{y_i \le Y < y_{i+1}\}$\\
%$1\le i\le m-1$ 
%%
%\end{flushleft}}
%& $P_i = p_0^i (1-p_0)$ & $N_i = (1-p_0) N$\\
%%\hline
%%---------------------------------------------------------
%$B_m = \{Y\ge  y_m\}$ & $P_m = p_0^m$ & $N_m = p_0 N$\\
%\hline
%%---------------------------------------------------------
%\end{tabular}
%\end{center}
%\end{table}

%--------------------------------------------------
\subsection{Remarks on related ideas}\label{sec:related}
%--------------------------------------------------
%--------------------
% my own work
%--------------------
Although the context is completely different, the key formula \eqref{eq:dF_intg} and the technique behind the derivation (Section~\ref{sec:theory}) were discovered after a recent work on analyzing the `failure mixing rate' in optimizing MCMC for rare events~\cite{Au2026_dR1}. 
%--------------
% theory
%--------------
On the other hand, compared with \eqref{eq:dF_surface} that is a surface integral over the (potentially high dimensional) input variable space ($\x$), \eqref{eq:dF_intg} is conceptually simpler as it is only a one-dimensional integral over the response gradient space.
The conditional expectation form in \eqref{eq:dF_E} allows one to develop probabilistic intuitions. It opens up possibilities for general data-analytics such as kernel regression; see the literature note at the end of Section~\ref{sec:kernel} later.
%--------------
% algorithm
%--------------
While this work was developed independently, on hindsight the resulting approximation via kernel smoothing with direct Monte Carlo samples is similar to \cite{Papaioannou2018} that was based on \eqref{eq:dF_surface}; see \eqref{eq:dF_mc} in this work and (11) in~\cite{Papaioannou2018}.
Nevertheless, the developed algorithms are different, i.e., based on SS in this work, and based on sequential importance sampling in~\cite{Papaioannou2018}.
As seen in \eqref{eq:dF_SS} and demonstrated in the numerical examples (Section~\ref{sec:examples}), the use of SS allows sensitivity calculations to be naturally embeded in existing codes of SS and to be obtained for all response threshold values ($y$) generated in an SS run.

%===============================================
\subsection{Organization of this work}\label{sec:organize}
%===============================================
The remaining parts of this work are organized as follows. 
Section~\ref{sec:theory} presents the theory and proves the key formula \eqref{eq:dF_intg}.
Section~\ref{sec:kernel} discusses Monte Carlo estimation via \eqref{eq:dF_intg} with kernel smoothing.
%
%Section~\ref{sec:eg01} illustrates the theory and computation using a simple problem where analytical solution for failure probability and sensitivity are available. 
%
Section~\ref{sec:examples} investigates the proposed methodology using a number of examples of different nature; see Table~\ref{tab:examples} for a summary.
%
%------------------------------------------------------
% sensitivity parameters of different nature
%------------------------------------------------------
Sensitivity parameters of special nature are illustrated, e.g., those with non-zero response gradient but zero sensitivity in $F$, such as $\hp_3$ in Example~1 (Normal response); those with a significant chance of zero response gradient, such as $\hp_2$ (second story stiffness) in Example~2 (shear building buckling); those displaying weak correlation between response and gradient, but still carrying significant sensitivity in $F$, such as $\hp_2$ (natural frequency) in Example~3 (first passage problem); and those in applications (Example~4, pile design) that may prefer finite difference over analytical means for response gradient to save coding effort. 
%

%================================================
\section{Theory of sensitivity and response gradient}\label{sec:theory}
%================================================
In this section we present the main theory, which shows that the sensitivity $\partial F/\partial \hp$ is given by \eqref{eq:dF_intg}.
The derivation assumes that the response $Y = f(\bX,\hp)$ and its gradient $\df= \partial f(\bX,\hp)/\partial \hp$ exist almost surely, and they have continuous joint, conditional and marginal PDFs. This is to justify their appearance in various places, and the Taylor approximation in~\eqref{eq:P'}.
Although the problem contexts are completely different, the technique used here is similar to that in Section~7.2 of a recent work \cite{Au2026_dR1} investigating the derivative of failure mixing rate w.r.t.\@ hyperparameters in optimizing MCMC for rare events. 

Technically, $\alpha$ affects $F(y,\hp) = P(Y\ge y)$ through $Y = f(\bX,\alpha)$, and hence any probability about $Y$.
The effect is implicit, and so direct differentiation of $F$ is not possible.
We will start from first principles, i.e., derivative as the limit of the ratio of perturbations. 
Consider a perturbation $\dhp$ in $\hp$. To the first order of $\Delta \hp$, $Y$ will be perturbed to
\begin{align}\label{eq:Yc'}
Y' 
= Y + \df \dhp
\end{align}
Replacing $Y$ in $F= P(Y\ge y)$ by $Y'$ and rearranging gives the perturbed $F$:
\begin{align}\label{eq:R'}
F'  = P(Y \ge y - \df \dhp)
\end{align}
Unlike the usual reasoning in Calculus, it is not legtimate to make the Taylor approximation $F'  \approx P(Y\ge y) - \df \dhp\, \partial P( Y \ge y)/\partial y$ because $\df$ is random and it is generally correlated with $Y$. 
In view of this, we condition on $\df = t$ to remove the randomness:
\begin{align}\label{eq:tmp03}
F' = \int P(Y \ge y - t \dhp|\df = t)\, p(\df=t)\, dt
\end{align}
For every $t$, we can now make the Taylor approximation for small $\dhp$:
\begin{align}\label{eq:P'}
&P(Y \ge y - t \dhp|\df=t)\nonumber\\
&= P(Y \ge y|\df=t)
- t\, \dhp \frac{\partial}{\partial y} P(Y \ge y|\df = t) + o(\dhp)\nonumber\\
&= P(Y \ge y|\df = t)
+ t\, \dhp\, p(Y=y|\df = t) + o(\dhp)
\end{align}
where $p(Y=y|\df=t)$ denotes the conditional PDF of $Y$ at $y$ given that $\{\df=t\}$.
Substituting \eqref{eq:P'} into \eqref{eq:tmp03} gives three integrals. The first is equal to $F$. Substracting it from LHS gives the perturbation $\Delta F = F' - F$. Dividing by $\Delta\hp$ and noting $p(Y=y|\df=t)\, p(\df=t) = p(\df=t,Y=y)$ gives 
\begin{align}
\frac{\Delta F}{\dhp} 
&= \int t\, p(\df=t,Y=y) dt + \frac{o(\dhp)}{\dhp}
\end{align} 
Taking $\Delta\alpha\rightarrow 0$ gives the required result in \eqref{eq:dF_intg}.

%================================================
\section{Monte Carlo with kernel smoothing}\label{sec:kernel}
%================================================
In this section we develop a Monte Carlo method for efficient calculation of reliability sensitivity, leveraging the key formula in \eqref{eq:dF_intg}.
A Monte Carlo approach is adopted to allow general applications, where only the point-wise value of the response function and its gradient are required, but otherwise no other information is needed. 
Subset Simulation (SS) is adopted for generating samples of frequent to rare events. 
As will be seen shortly, the method allows one to calculate $\partial F(y,\hp)/\partial \hp$ in the same SS run for $F(y,\hp)/\partial \hp$, covering all values of $y$ generated in the SS run from high to low probability.

Equation \eqref{eq:dF_intg}, or its conditional expectation form in \eqref{eq:dF_E}, suggests averaging over Monte Carlo samples  of $G$. 
A fundamental difficulty arises from the conditioning on $\{Y=y\}$, which is a zero-probability event. In typical situations, Monte Carlo samples of $G$ are associated with random values of $Y$, and so it is impossible to obtain samples with $Y=y$ exactly. 
In this work we overcome this difficulty using the concept of `kernel smoothing'. It allows one to use samples in the neighborhood of $\{Y=y\}$ for Monte Carlo averaging at the expense of introducing a bias, which is nevertheless controllable.   
In what follows, we first introduce the idea and discuss how the integral in \eqref{eq:dF_intg} can be estimated by direct Monte Carlo, i.e., using samples of $\{Y,\df\}$ where $\bX\sim q$. 
After that we extend the idea to SS, where the samples are conditional on different threshold bins of $Y$. 
Kernel smoothing is not new and is a topic in itself in data-science. See a literature note near the end of Section~\ref{sec:kernel_width} later. 

%===================================
\subsection{Direct Monte Carlo}\label{sec:kernel_mc}
%===================================
The basic idea of kernel smoothing in our context is to approximate the integral in \eqref{eq:dF_intg} by a double integral over $t$ and $y$ so that it can be estimated by Monte Carlo averaging over samples of $\{\df,Y\}$. To conform with the condition $\{Y=y\}$,  a weighting is introduced so that only contributions of the samples in the neighborhood of $\{Y=y\}$ are accounted for. 
Specifically, let $K(\tau)$ be a user-defined univariate PDF. For its role in the problem that will become clear shortly, it is called a `kernel function', and is often chosen to have a centralized shape around zero and significant probability over a length scale of $O(1)$, e.g., standard Normal PDF. 
It can be easily verified that for given $w>0$ and $y\in \real$, a function of $\tau$ in the form $w^{-1} K\left((\tau-y)/w\right)$ is still a univariate PDF, but now it is centered at $y$ and varies over a scale of $w$. 
Consider now an integral of the product of sensitivity and kernel:
\begin{align}\label{eq:J_mcs}
\begin{split}
J(y,\hp) 
&= \int \frac{\partial F(\tau,\hp)}{\partial \hp} w^{-1} K\left(\frac{\tau-y}{w}\right) d\tau\\
&= \iint t\, w^{-1} K\left( \frac{\tau-y}{w}\right)\, p(G=t,Y=\tau)\,dt\,d\tau\\
&= E\left[G\, w^{-1} K\left( \frac{Y-y}{w}\right)\right]
\end{split}
\end{align}
where the expectation is taken with $\{G,Y\}$ distributed as $p(G=t,Y=y)$; the second equality is obtained by substituting \eqref{eq:dF_intg}. 
Intuitively, from the first expression in \eqref{eq:J_mcs}, $J(y,\hp)$ collects the contributions of $\partial F (\tau,\hp)/\partial\hp$ over a neighorhood of width $w$ around $\tau = y$. 
As $w\rightarrow 0$, the function $w^{-1} K\left((\tau-y)/w\right)$ tends to a Dirac Delta function centered at $\tau=y$. One can then expect that $J(y,\hp)\rightarrow \partial F(y,\hp)/\partial \hp$. 
This is the idea of kernel smoothing in our context, where $J$ is taken as a proxy for $\partial F/\partial \hp$. The merit is that it can be estimated by averaging the term inside $E[\cdot]$ over Monte Carlo samples of $\{G,Y\}$, which is feasible now because the zero-probability condition $\{Y=y\}$ has been removed. 
See the resulting estimator in \eqref{eq:dF_mc}. 
Of course, this feasibility is made at the expense of a bias, but which can be suppressed with sufficiently small $w$.  
In particular, if $K(\cdot)$ is symmetric about $0$ and $\partial F(\tau,\hp)/\partial \hp$ is twice differentiable w.r.t.\@ $\tau$ at $\tau=y$, then the bias is second order, i.e., 
\begin{align}\label{eq:bias}
J(y,\hp) = \frac{\partial F}{\partial \hp} + O(w^2)
\end{align}
This can be reasoned by substituting the Taylor series 
\begin{align}
\frac{\partial F(\tau,\hp)}{\partial \hp} =
 \frac{\partial F(y,\hp)}{\partial \hp} + \frac{\partial^2 F(y,\hp)}{\partial y \partial \hp} (\tau-y) + O\left( (\tau-y)^2 \right)
\end{align}
into the first expresion in \eqref{eq:J_mcs} and noting that upon integration the zeroth order term gives the target $\partial F(y,\hp)/\partial \hp$, the first order term gives zero, and the reminder (second order) term gives $O(w^2)$.
The choice of $w$ should balance between estimation bias and variance. See Section~\ref{sec:kernel_width} later.

%===================================
\subsection{Subset Simulation}\label{sec:kernel_ss}
%===================================
The idea of kernel smoothing in the last section has resolved the difficulty of zero-probability conditioning, but issues with rare events still remain. Unless the number of samples is very large ($\propto 1/F$), direct Monte Carlo does not offer enough samples (if any) of $\{\df,Y\}$ for estimating small $F$ or its sensitivity $\partial F/\partial \hp$. 
In this work we use Subset Simulation (SS) to address rare events, generating samples covering frequent to small regimes of the CCDF of $Y$. 
As the samples in SS are conditional on a series of thresholds, their occurence needs to be properly weighted in the Monte Carlo average. 
Below we show how that can be done to give a proper estimator. 

Consider a conventional SS run where $N$ samples of $\bX$ and $Y = f(\bX,\hp)$ are generated at each simulation level $i$ ($=0,1,\dots,m-1$) conditional on $\{Y\ge b_i\}$, with $b_0=-\infty$, and $b_1<b_2<\cdots<b_m$ an increasing sequence of generated thresholds such that on a sample basis $P(Y\ge b_{i+1}|Y\ge b_i) = p_0$, i.e., the `level probability'.  
For sensitivity calculation, the theory in this work requires that the response gradient $G = \partial f(\bX,\hp)/\partial \hp$ be also calculated for all samples of $\bX$.  
Let the samples be grouped into a series of `threshold bins' $\{B_i\}_{i=0}^{m-1}$ that partition the real line. Table~\ref{tab:bins} summarizes the definition of the bins $B_i$, their probability $P_i$ and the number of samples $N_i$ they contain.  
Using the theorem of total probability, \eqref{eq:J_mcs} can be written as
\begin{align}\label{eq:J_ss}
J(y,\hp) 
= \sum_{i=0}^{m-1} E\left[G\, w^{-1} K\left( \frac{Y-y}{w}\right) \bigg| B_i\right] P(B_i)
\end{align}
The conditional expectation $E[\cdot|B_i]$ can be estimated by averaging over the samples in $B_i$: 
\begin{align}\label{eq:Ebin}
E\left[G\, w^{-1} K\left( \frac{Y-y}{w}\right) \bigg| B_i\right]
\approx N_i^{-1} \sum_{k=1}^{N_i} G_{ik} w^{-1} K\left( \frac{Y_{ik}-y}{w}\right)
\end{align}
where the subscript $ik$ denotes the $k$th sample of $B_i$. 
Although it has not be explicitly indicated, the kernel width $w$ can depend on $y$, i.e., `adaptive', to account for the density of $\{Y_{ik}\}$ around $y$ or other effects such as relationship with $G$.
Substituting \eqref{eq:Ebin} into \eqref{eq:J_ss} and writing $P(B_i) = P_i$ gives the key estimation formula in \eqref{eq:dF_SS}.
In theory the threshold $y$ can be any value, although for proper estimation it should be limited to the range of values generated in an SS run. 
Taking computer coding into consideration, a convenient way is to calculate the sensitivity at the generated thresholds of $Y$. In this manner, an SS run gives an estimate of the CCDF curve $F(y,\hp)$ (as conventional), as well as the `sensitivity curve' $\partial F(y,\hp)/\partial \hp$ (new in this work) for all generated values of $y$.

%--------------------------------------
% applicable for any MCMC
%--------------------------------------
% spare ref:
%  Neal2011 on HMC
As long as the conditional samples in different bins are available, the methodology here is  applicable regardless of how they are generated in SS, e.g., independent component Metropolis random walk in the debut paper~\cite{Au2001_subsim}, conditional sampling~\cite{Papaioannou2015} that is an elegant finite-infinite dimensional equivalent~\cite{Au2016}, Hamiltonian Monte Carlo~\cite{Duane1987,Wang2019} or its Riemannian variant~\cite{Girolami2011,Chen2022}, and affine transformation~\cite{Goodman2010,Shields2021}.  
%
% spare ref:
%  nuclear fuels~\cite{Dhulipala2022}
%  hydrogen storage tanks~\cite{SidAmer2024}
%  air traffic~\cite{Beller2024}
%  maritime structures~\cite{Lima2023}
Aside, see recent applications of SS to autonomous driving~\cite{Tian2024}, climate simulation~\cite{Finkel2024} and fusion energy systems~\cite{Simon2025}.

%===================================
\subsection{Kernel width}\label{sec:kernel_width}
%===================================
As mentioned before, the key idea in kernel smoothing that resolves the issue of zero-probabilty conditioning $\{Y=y\}$ is the use of kernel $K(\cdot)$ that allows averaging over Monte Carlo samples of $\{Y\ne y\}$ but at the same time weighting them so that only those sufficiently close to $\{Y=y\}$ matters.
The weights are controlled by the kernel width $w$, which should be chosen to balance over estimation bias and variance.  
A small $w$ can reduce bias because only samples near $\{Y=y\}$ receive a significant weight, but it will increase the variance of Monte Carlo average because the effective number of samples is reduced; and vice-versa. 
In between the extremes of small-bias-high-variance and high-bias-low-variance lies the optimal $w$.
Intuitively, the optimal $w$ depends on the total number of samples. For example, more samples means lower variance and hence can accomodate a smaller $w$. 

%-----------------
% Scott's rule
%-----------------
Similar to many algorithm parameters, the choice of kernel width is an open issue that depends on the distribution of data but which is not known a priori. 
%
%See Section~\ref{sec:kernel_review} later for a literature note. 
%
In this work we adopt the Scott's rule~\cite{Scott1979,Scott1992} for its simplicity and theoretical elegance (optimal for Normal data). For $y\in B_i$ (Table~\ref{tab:bins}), the kernel width is taken as 
\begin{align}\label{eq:scott}
w_i = \sigma_Y \left(\frac{4}{3N_i}\right)^{1/5}
\end{align}
where $N_i$ is the number of samples in $B_i$, and  
\begin{align}\label{eq:sigma_y}
\sigma_Y^2 = E[Y^2] - E[Y]^2 
\end{align}
is the variance of $Y$ when $\bX\sim q$. The statistical moments can be estimated in a single SS run using the samples in all bins, by virtue of the theorem of total probability ($r=1,2$):
\begin{align}
\label{eq:moment_y}
E[Y^r] = \sum_{i=0}^{m-1} E[Y^r|B_i] P_i
=
\sum_{i=0}^{m-1} P_i N_i^{-1}\sum_{k=1}^{N_i} Y_{ik}^r
\end{align}

While the specific formula and relevant statistic of the optimal kernel width depends on the criterion and distribution of samples, the scaling $w_i\propto N_i^{-1/5}$ has an origin that can be reasoned intuitively as follows.
A Monte Carlo average with kernel smoothing of width $w$ is effectively an average of $w\,N_s$ samples, where $N_s$ is the effective number of independent samples. 
Its variance is therefore $O(w^{-1}N_s^{-1})$.
If the bias is $O(w^p)$ for some $p$ (say), the mean squared error (MSE) of the average from the target is of the form $a\,w^{-1}N_s^{-1} + b\,w^{2p}$, where $a$ and $b$ are constants that depend on the distribution and location of estimate ($y$). 
Minimizing the MSE gives the optimal width of $w = \left( a/2bpN \right)^{1/(2p + 1)}$. Taking $p = 2$ for second order bias (see \eqref{eq:bias}) then gives $w \propto N_s^{-1/5}$ as implied by \eqref{eq:scott}.
Intuitively, a larger $N_s$ allows one to reduce the bias by narrowing $w$, but the latter should proceed at a slower rate than $O(N_s^{-1})$ to suppress the growth of variance due to the lack of samples in the $O(w)$ neighborhood of $\{Y=y\}$.

As a literature note, kernel smoothing is a nonparametric technique for estimation involving conditional expectation given one or more features by constructing a weighting, i.e., kernel function. 
See \cite{Nadaraya1964} an early work by Nadaraya, and \cite{Bierens1994} a monograph in econometrics. 
Besides the kernel, performance also depends on the locally weighted regression forms, e.g., constant, linear or polynomial.
The form in this work is a simple conventional choice often referred as the Nadaraya-Watson model, which corresponds to a constant locally weighted regression form. 
See Chapter~5 of \cite{Wand1994} that discusses general `local polynomial estimators', \cite{Atkeson1997} a review, and~\cite{ek2020} a general review on nonparametric regression.

\section{Illustrative examples}\label{sec:examples}
%===============================================
In this section we present a series of examples to illustrate the theory and computational methodology. 
As is conventional in SS, the input random variables $\{X_i\}_{i=1}^n$ are assumed to be i.i.d.\@ N(0,1). 
In each example we first present the setup, how to calculate the response $Y = f(\bX,\hp)$ and its gradient $\partial Y/\partial \hp$ w.r.t.\@ sensitivity parameter $\hp$. If available, analytical solution for the failure probability $F = P(Y\ge y)$ and its sensitivity $\partial F/\partial \hp$ will also be presented. 
In the numerical investigation, $F$ and $\partial F/\partial \hp$ will be calculated using SS in the same Monte Carlo run.  
The results of sensitivity will be compared with a benchmark. The latter is taken as the analytical solution when available (Examples 1 and 2), or otherwise as the central difference of direct Monte Carlo estimates with common random numbers (Examples 3 and 4).
Results of a single SS run will be presented to give an idea of typical results. 
Statistics from 1000 independent runs will also be presented to assess the bias and variance of sensitivity estimates. 
%

%-------------------------------------------
% implementation details of SS
%-------------------------------------------
SS will be implemented for $m = 3$ levels with a level probability of $p_0 = 0.1$ and $N = 1000$ samples per level. This is intended to provide samples populating the CCDF of $Y$ from 1 down to at least $p_0^3 = 10^{-3}$. There will be $m=3$ bins, i.e., $B_0$ and $B_1$ with $(1-p_0)N = 900$ samples, and $B_3$ with $N=1000$ samples. 
Conditional sampling~\cite{Papaioannou2015} is used for generating MCMC samples. In level $i\ge 1$ the candidate (before checking $Y\ge b_i$) is generated as $\bX' = a_i \bX + s_i \bZ$ where $\bX$ is the current sample, $\bZ$ is an i.i.d.\@ N(0,1)$^n$ vector independent of $\bX$, $a_i\in (0,1)$ is a correlation (hyper-) parameter and $s_i = \sqrt{1-a_i^2}$. 
We take $a_i = 0.5(1+u_i/v_i)$ where $u_i = \PPhi^{-1}(p_0^i)$ and $v_i = \PPhi^{-1}(p_0^{i+1})$; $\PPhi(\cdot)$ denotes the standard Normal CCDF. This choice minimizes the correlation between successive MCMC samples of Normal response, and is found to be a good for other applications~\cite{Au2025}.

Four examples serving complementary purposes are considered. See Table~\ref{tab:examples} for a summary. 
%------------------------------------------
% example 1: Normal response
%------------------------------------------
The first example (Section~\ref{sec:eg01}) considers Normal response with a simple relationship betwen input variables and output response, providing a clear exposition of theory with analytical solution for benchmarking.  
All sensitivity parameters have well-behaved response gradients.
%
% special parameter
%
Parameter $\hp_3$ is designed to illustrate the possibility of non-zero response gradient but zero reliability sensitivity, prompting for care in applying intuitions and interpreting sensitivity results in proper scales.

%----------------------------
% example 2: buckle
%----------------------------
The second example (Section~\ref{sec:buckle}) considers the global buckling of a shear building. It serves to outline the basic ingredients for general structures while admiting analytical solution for the special case of shear building for benchmarking. 
The example is scalable with the number of stories and input random variables, although for illustration a five-storied building with five i.i.d.\@ Lognormal random variables for the floor loads is considered. 
The mean value of random floor loads ($\hp_1$) and the deterministic stiffness of the second story ($\hp_2$) are considered to illustrate different types of sensitivity parameters. 
%
% special parameter
%
The latter is designed to illustrate the case when the response gradient has a significant chance to be zero. This is typical in systems with components connected in series, where system failure is insensitive to components that have a relatively low failure rate than others. 
As will be seen, this renders the influence of the parameter a `rare event' itself, inflating the variance of sensitivity estimate and calling for advanced strategies for variance reduction.

%----------------------------
% example 3: SDOF
%----------------------------
The third example (Section~\ref{sec:sdof}) considers the first passage failure of a single-degree-of-freedom (SDOF) structure subjected to white noise excitation. 
Despite the SDOF nature, the problem is complicated for its large number of input random variables ($n=400$) in modeling the excitation, the same large number of failure modes induced by the possible peak response time, and dynamics that induces non-trivial correlation among them. The problem is representative of a high dimensional problem, and has been considered in previous works investigating MCMC for rare events~\cite{Au2025,Au2026_dR1}. 
See also a review on first passage problems~\cite{Jiang2025}.
%
%Analytical solution for failure probability or sensitivity is not available. 
%
%For the linear-elastic nature of response, a very efficient importance sampling method is avaiable for the failure probability~\cite{Au2001_isee}, 
%
As analytical solution for failure probability or sensitivity is not available, central difference with direct Monte Carlo is used for benchmarking.   
Sensitivities to the damping ratio ($\hp_1$) and natural frequenty ($\hp_2$) are considered, which are of very different nature.
%
% special parameter
%
The latter is designed to illustrate the possibility of a parameter displaying weak correlation between the response and its gradient, but still carrying signficant reliability sensitivty.  As will be seen, the resulting sensitivity estimate can have large variance, calling for advanced strategies for variance reduction.

%----------------------------
% example 4: geotech
%----------------------------
The fourth example (Section~\ref{sec:pile}) considers the risk of excessive settlement in the serviceability limit state design of pile foundation.
The problem is considered for its relevance in application, and complication in its spatially varying soil friction angle along the depth. The latter is modeled by $n=120$ input random variables transformed through Cholesky decomposition of correlation matrix to produce a Lognormal random field with specified mean, standard deviation and correlation length. 
Sensitivities to the pile diameter ($\hp_1$) and the mean of friction angle ($\hp_2$) are considered, whose results are intuitive but still non-trivial especially in the correlation between response and its gradient. 
Different from other examples, the response gradient is calculated using finite difference, which is a scenario in applications where only response function values can be accessed, or it is preferred over analytical means to simplify coding effort. 
Similar to Example~3, analytical solution for sensitivity is not available, and so central difference with direct Monte Carlo is used for benchmarking.

\begin{center}\footnotesize
\begin{longtable}{m{1.5em} m{1.5em} m{12em} m{8em} m{11em}}
\caption{Summary of examples. Input random variables $\{X_i\}_{i=1}^n$ i.i.d.\@ N(0,1). (Output) response $Y = f(\bX,\hp_i)$. $\hp_i = $ sensitivity parameter. Failure probability $F = P(Y\ge y)$, sensitivity $\partial F/\partial \hp_i$}\label{tab:examples}
\\
%---------------------------------------------------------
\hline
No. & $n$ & $Y$ & $\partial Y/\partial \hp$ & Remarks\\
\hline
%---------------------------------------------------------
1 & 2 & 
\vbox{\begin{flushleft}
Normal response\\
$Y = \hp_1 + (\hp_2^2-\hp_3)^{1/2} X_1 + \hp_3 X_2$
\end{flushleft}}
& \eqref{eq:eg01_dY12} and \eqref{eq:eg01_dY3} & 
\vbox{\begin{flushleft}
$\hp_3$ has $\partial Y/\partial \hp_3\ne 0$ but $\partial F/\partial \hp_3=0$\\
\hfill\\
Analytical solution see \eqref{eq:eg01_exact_F} and \eqref{eq:eg01_exact_dF}
\end{flushleft}}
\\
\hline\\
%---------------------------------------------------------
2 & 5 &
\vbox{\begin{flushleft}
Shear building buckling\\
$Y = \lambda_0/\lambda$\\
$\lambda = $ from eigenvalue problem \eqref{eq:buckle_eig}\\
$\lambda_0 = $ normalizing factor\\
\hfill\\
Lognormal vertical loads $W_i = w\, e^{a+b X_i}$\\
$a$ and $b$ from \eqref{eq:root}\\
\hfill\\ 
$\hp_1 \equiv w$\\
$\hp_2 \equiv $ 2nd story stiffness
\end{flushleft}}
&
\eqref{eq:buckle_dY}, solve matrix equation \eqref{eq:buckle_deriv}
& 
\vbox{\begin{flushleft}
$\hp_2$ is difficult, with 80\% chance of $\partial Y/\partial \hp_2=0$\\
\hfill\\
Analytical solution see \eqref{eq:buckle_F_num}, \eqref{eq:buckle_dF_num01} and \eqref{eq:buckle_dF_num02}
\end{flushleft}}
\\
\hline\\
%---------------------------------------------------------
3 & 400 & 
\vbox{\begin{flushleft}
Linear SDOF first passage\\
$Y = \max_{j=0}^{n-1} |u(j\Delta t)|$\\
$\ddot{u}+2\zeta\omega\dot{u}+\omega^2 u = W$\\
$\omega=2\pi$ (rad/s), $\zeta=1\%$\\
$u(0) = \dot{u}(0) = 0$\\
\hfill\\
$W_i = (\frac{2\pi S}{\Delta t})^{1/2} X_{i+1}$\\
$0\le i\le n-1$\\
$\Delta t = 0.05$~s, $S = 0.86$~N$^2$/(rad/s)\\
\hfill\\
$\hp_1 \equiv \zeta, \hp_2 \equiv \omega$
\end{flushleft}}
&
\eqref{eq:sdof_dY}, solve ODE \eqref{eq:sdof_du01} and \eqref{eq:sdof_du02}
&
\vbox{\begin{flushleft}
$\hp_2$ is difficult, apparently weak correlation between $Y$ and $\partial Y/\partial \hp_2$, but still signficant sensitivity $\partial F/\partial \hp_2$\\
\hfill\\
No analytical solution\\
Benchmark with central difference of $F$ from direct Monte Carlo. Relative step size 1\%
\end{flushleft}}
\\
\hline
%---------------------------------------------------------
4 & 120 &
\vbox{\begin{flushleft}
Pile serviceability design\\
$Y = F_{50}/Q_{sls}$\\
$F_{50} =$~design load (fixed)\\
$Q_{sls} = $ effective soil resistance (random)\\
See \eqref{eq:Qsls} and Table~\ref{tab:pile}\\
\hfill\\
Soil friction angles $\{\phi'_i\}_{i=1}^n\sim $ Lognormal with\\
mean $\mu = 0.5585$~rad (32$^\circ$), 
c.o.v.\@ = 17\%, and
correlation length $4$~m\\
\hfill\\
$\hp_1 \equiv B$ (pile diameter)\\
$\hp_2 \equiv \mu$ 
\end{flushleft}}
& Central difference with relative step size 0.1\% & 
\vbox{\begin{flushleft}
No analytical solution\\
Benchmark with central difference of $F$ from direct Monte Carlo. Relative step size 1\%
\end{flushleft}}
\\
\hline
%
%---------------------------------------------------------
\end{longtable}
\end{center}
%\end{table}

%##################
\FloatBarrier
%##################

%===============================================
\subsection{Normal response}\label{sec:eg01}
%===============================================
%Consider the response $Y = W_1 + W_2$ where $W_1\sim N(\hp_1,\hp_2^2)$, i.e., Normally distributed with mean $\hp_1$ and variance $\hp_2^2$, and $W_2\sim N(0,1)$; $W_1$ and $W_2$ are jointly Normal with covariance $\hp_3\in (0,1)$.  
%%
%We are interested in the sensitivities of $P(Y\ge y)$ w.r.t.\@ the parameters in $\hpb = \{\hp_1,\hp_2,\hp_3\}$. 
%%
%If we define $\{W_1,W_2\}$ as input random variables then $\hpb$ affects the PDF of $\{W_1,W_2\}$, which violates the assumption in this work that their PDF does not depend on the parameters. This is a typical situation encountered in applications. 
%%
%To fit the context, we can reparameterize $W_1 = \hp_1 + (\hp_2^2-\hp_3^2)^{1/2} X_1 + \hp_3 X_2$ and $W_2 = X_2$, where $\{X_1,X_2\}$ are i.i.d.\@ $N(0,1)$ and are taken as input random variables instead of $\{W_1,W_2\}$.
%%
%Check that the distribution is the same as before, i.e., $W_1\sim N(\hp_1,\hp_2^2)$ and $W_2\sim N(0,1)$, and their covariance is still $\hp_3$.
%%
Consider the response defined by 
\begin{align}\label{eq:eg01_Y}
Y = \hp_1 + (\hp_2^2 - \hp_3^2)^{1/2} X_1 + \hp_3 X_2
\end{align} 
where $\{\hp_1,\hp_2,\hp_3\}$ are sensitivity parameters and $\{X_1,X_2\}$ are i.i.d.\@ N(0,1). 
%
%We are interested in the sensitivities of $P(Y\ge y)$ w.r.t.\@ $\{\hp_1,\hp_2,\hp_3\}$. 
%
It can be reasoned that $Y\sim N(\hp_1,\hp_2^2)$ and so
{
\begin{align}\label{eq:eg01_exact_F}
F(y,\hpb) = P(Y\ge y) = \Phi(-z)
\qquad z = \frac{y-\hp_1}{\hp_2}
\end{align}
}
where $\Phi(\cdot)$ denotes the standard Normal CDF.
Taking derivatives gives
{
\begin{align}\label{eq:eg01_exact_dF}
\frac{\partial F}{\partial \hp_1} = \hp_2^{-1}\phi(z)
\qquad
\frac{\partial F}{\partial \hp_2} = \hp_2^{-1} z\,\phi(z)
\qquad
\frac{\partial F}{\partial \hp_3} = 0
\end{align}
}

%--------------------------------------------------------------------------------------
\subsubsection{Analytical verification with theory}\label{sec:eg01_anal}
%--------------------------------------------------------------------------------------
We now determine the sensitivities using the theory in this work, i.e., \eqref{eq:dF_E}, and verify that they are the same as those in \eqref{eq:eg01_exact_dF}. The basic steps are 1) derive $\partial Y/\partial \hp_i$, 2) determine the conditional distribution and hence expectation of $\partial Y/\partial \hp_i$ given $\{Y=y\}$, and 3) substitute all results into \eqref{eq:dF_E}.  
Direct differentiation of \eqref{eq:eg01_Y} gives
\begin{gather}\label{eq:eg01_dY12}
\frac{\partial Y}{\partial \hp_1} = 1
\qquad
\frac{\partial Y}{\partial \hp_2} = \hp_2 (\hp_2^2-\hp_3^2)^{-1/2} X_1
\\
\frac{\partial Y}{\partial \hp_3} = -\hp_3 (\hp_2^2-\hp_3^2)^{-1/2} X_1 + X_2
\label{eq:eg01_dY3}
\end{gather}
%---------------------
% dF/da1
%---------------------
Since $\partial Y/\partial \hp_1$ is a constant, it is clear that $E[\partial Y/\partial \hp_1|Y=y] = 1$. 
%---------------------
% dF/da2
%---------------------
%
For $E[\partial Y/\partial \hp_2|Y=y]$, we need to determine $E[X_1|Y=y]$. 
Note from \eqref{eq:eg01_Y} that $Y\sim N(\hp_1,\hp_2^2)$ is jointly Normal with $X_1\sim N(0,1)$ with covariance $(\hp_2^2-\hp_3^2)^{1/2}$. 
Recall from standard results that if $Z_1\sim N(\mu_1,\sigma_1^2)$ and $Z_2\sim N(\mu_2,\sigma_2^2)$ are jointly Normal with covariance $c$, then $E[Z_2|Z_1] = (Z_1-\mu_1)\, c/\sigma_1^2$. Applying this result gives
\begin{align}
E[X_1|Y=y] &= \hp_2^{-2} (\hp_2^2-\hp_3^2)^{1/2} (y-\hp_1)
\end{align}
Multiplying this with the PDF of $Y$, i.e., $\hp_2^{-1} \phi(z)$ where $z = (y-\hp_2)/\hp_3$, gives the same expression for $\partial F/\partial \hp_2$ in \eqref{eq:eg01_exact_dF}. 
%---------------------
% dF/da3
%---------------------
Finally, the benchmark result $\partial F/\partial \hp_3 = 0$ in \eqref{eq:eg01_exact_dF} appears counter-intuitive, as $\partial Y/\partial \hp_3$ in \eqref{eq:eg01_dY3} is generally non-zero. 
Equation~\eqref{eq:dF_E} does give the same result, however. To see this, evaluating the (unconditional) covariance of $\{Y,\partial Y/\partial \hp_3\}$ using $Y$ in \eqref{eq:eg01_Y} and $\partial Y/\partial \hp_3$ in \eqref{eq:eg01_dY3} shows that it is zero. Since $Y$ and $\partial Y/\partial \hp_3$ are jointly Normal, this implies that are independent and so $E[\partial Y/\partial \hp_3|Y=y] = E[\partial Y/\partial \hp_3] = 0$.
 
%---------------------------------------------------------------------------
\subsubsection{Single SS run}\label{sec:eg01_single}
\begin{figure}[h]
{
\includegraphics[width=\textwidth]{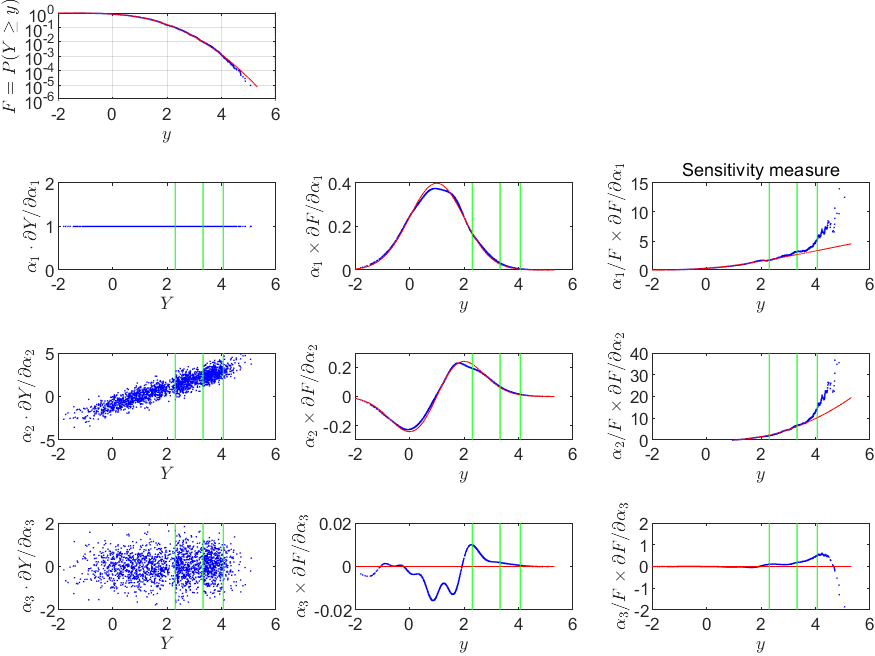}
\caption{Example 1 (Normal response). Results of a typical SS run with $m = 3$ levels, $p_0 = 0.1$ and $N=1000$ samples per level. Blue = SS, red = Benchmark, vertical green lines = thresholds at $\{10^{-1}, 10^{-2}, 10^{-3}\}$. The right column shows the fractional change of $F$ per fractional change of sensitivity parameter, the recommended sensitivity measure.}
\label{fig:eg01_single}
}
\end{figure}

%---------------------------------------------------------------------------
For illustration, consider $\hp_1 = \hp_2 = 1$ and $\hp_3 = 0.5$. 
Figure~\ref{fig:eg01_single} shows the results of a typical SS run. 
In all plots, the blue dots and red lines indicate the results of SS and benchmark, respectively.
%--------------------------
% vertical green lines
%--------------------------
The vertical green lines mark the threshold levels $\{y_1,y_2,y_3\}$ at CCDF values $\{p_0^i\}_{i=1}^m = \{10^{-1}, 10^{-2}, 10^{-3}\}$. 
Counting from the left to right, the $(1-p_0)N = 900$ samples to the left of the first vertical line go to $B_0$; those (same number) between the first and second line go to $B_1$; the remaining $N = 1000$ samples go to $B_2$. 
The quality of estimates for $F$ and $\partial F/\partial \hp$ to the right of the third line is deteriorating fast as it runs out of samples there.
%  

%==============================================
%----------------------------------------------------------
% eg01 - single run
%----------------------------------------------------------
%==============================================
%##################
\FloatBarrier
%##################

%------------------
% top-left: CCDF
%------------------
The top-left plot shows the CCDF estimate of $Y$, the basic output of SS that is commonly reported. 
%--------------------------
% left: scatter of dy vs y
%--------------------------
%
The rows below the top show the results that illustrate the theory and computation of this work, one row for each sensitivity parameter, and logically from left to right.
Consider for example the second row, i.e., $\hp_1$. 
The left plot shows the samples of response gradient (normalized) versus the response values in all bins. The former has been multiplied with the sensitivity parameter ($\hp_1$) so that it has the same unit as the response.
The response gradient normalized this way can be interpreted as the rate of change of $Y$ per fractional change of sensitivity parameter.
Consistent with the first expression in \eqref{eq:eg01_dY12}, the sample values of $\partial Y/\partial \hp_1$ are all equal to 1.

%--------------------------
% middle: dF vs y
%--------------------------
The middle column in Figure~\ref{fig:eg01_single} shows the sensitivity estimate (normalized) using SS and kernel smoothing, i.e., \eqref{eq:dF_SS}. 
%
%A constant kernel width (standad deviation) based on Scott's rule is used, where the variance of response is estimated using the samples in all bins; see \eqref{eq:sigma_y} and \eqref{eq:moment_y}.
%
It is dimensionless, and can be interpreted as the rate of change of $F$ per fractional change of the sensitivity parameter. 
For $\hp_1$, the sensitivity takes the shape of a Normal PDF. This is consistent with the exact solution in the first expression of \eqref{eq:eg01_exact_dF}.
The SS (blue) estimate generally agrees with the benchmark (red). It populates over the central as well as the upper tail (low probability) region.
The sensitivity tends to zero for large $y$ because the PDF of $Y$ does. This is typical and can be explained by \eqref{eq:dF_E}, where $\partial F/\partial \hp = p(Y=y)\, E[G|Y=y] \rightarrow 0$ if $E[G|Y=y]<\infty$.  
Of course, this trivial diminution does not imply that the parameter is no longer sensitivity for large $y$. It is just a matter of scaling and perspective, similar to the trivial convergence to zero of the variance $p(1-p)/N$ of direct Monte Carlo estimate for failure probability $p\rightarrow 0$.

%--------------------------
% middle: dF/F vs y
%--------------------------
In view of the trivial deminution, a more useful measure is to further divide the sensitivity by the CCDF value $F$. This leads to the sensitivity measure in the right column of Figure~\ref{fig:eg01_single}, which can be interpreted as the fractional change of $F$ due to fractional change in the sensitivity parameter.   
The PDF and hence diminution effect has been removed. 
On a fractional basis, the sensitivity w.r.t.\@ $\hp_1$ indeed increases as $y$ increases.  
For example, at $F\approx 10^{-3}$ ($y\approx 4$), a 1\% change in $\hp_1$ leads to about 3\% change in $F$. The same 1\% change in $\hp_2$ leads to 10\% change in $F$. In this sense $F$ is 3 times more sensitive to $\hp_2$ than $\hp_1$. 

%----------------------------
% alpha 2
%----------------------------
The third row in Figure~\ref{fig:eg01_single} shows the results for $\hp_2$. They are qualitatively similar to those of $\hp_2$, except that the response gradient (left plot) is now correlated with the response, and the sensitivity (middle plot) can be negative; see the second expression in \eqref{eq:eg01_exact_dF} and \eqref{eq:eg01_dY12}.
% 
%----------------------------
% alpha 2
%----------------------------
The results for $\hp_3$ in the bottom row are qualitatively different from those for $\hp_1$ or $\hp_2$. 
Recall from Section~\ref{sec:eg01_anal} that $\hp_3$ does not affect $F$ (hence zero sensitivity) but $\partial Y/\partial \hp_3$ is not zero. 
The scatter plot on the left displays a lack of correlation, not surprising because $\partial Y/\partial \hp_3$ and $Y$ are indeed uncorrelated. 
Despite this, the sensitivity estimate in the middle or right plot is non-zero, simply because of estimation error.  
This reflects the importance of viewing sensitivity estimates with the right scaling and order of magnitude in perspective.
For example, the oscillatory behavior with $y$ in the middle plot is merely an estimation error, which can be misleading when the order of magnitude is not paid attention to.
The sensitivity estimate on the right plot is not zero because of estimation error, but it has a small magnitude ($<1\%$) compared to other parameters, delivering the correct message of low sensitivity. 
The divergence beyond the third vertical line is expected, as it runs out of samples there.

%---------------------------------------------------------------------------
\subsubsection{Bias and variance from 1000 SS runs}\label{sec:eg01_stats}
\begin{figure}[h]
{
\includegraphics[width=\textwidth]{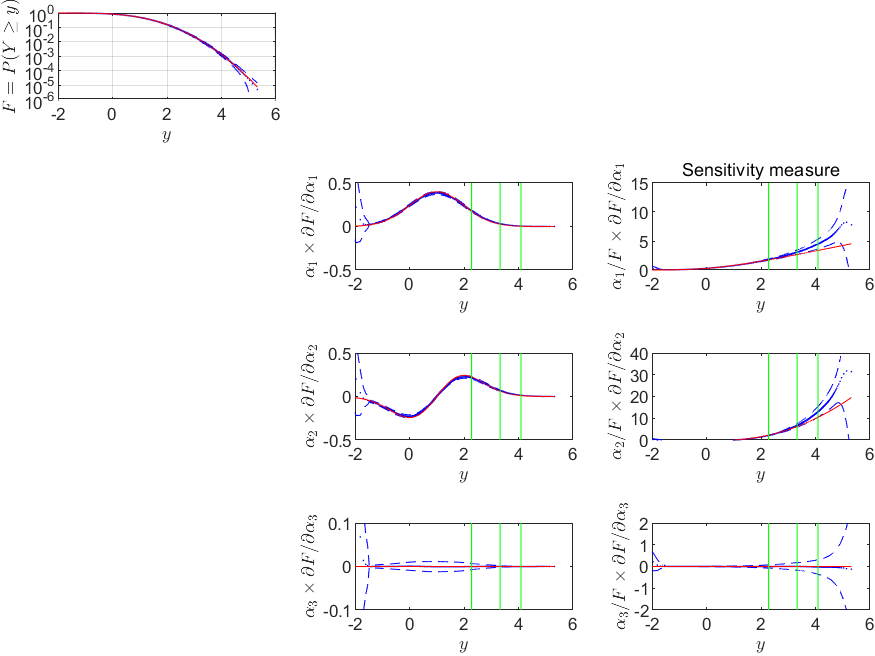}
\caption{Example 1 (Normal response). Sample mean (blue solid line) and $\pm 1\sigma $ bounds (blue dashed line) from 1000 SS runs. Red = Benchmark. Vertical green lines = thresholds at $\{10^{-1}, 10^{-2}, 10^{-3}\}$ of SS sample mean. Same layout as Figure~\ref{fig:eg01_single} except that scatter plots in the left column are omitted.}
\label{fig:eg01_stats}
}
\end{figure}

%---------------------------------------------------------------------------
Figure~\ref{fig:eg01_stats} shows the sample mean and $\pm 1 \sigma$ bound of the sensitivity estimate from 1000 i.i.d.\@ SS runs. 
It has the same layout as Figure~\ref{fig:eg01_single} except that the scatter plots in the left column are omitted, as they are no longer relevant. 
%----------------------------------------------
% top-left: CDF unbiased, no suprise
%----------------------------------------------
The top-left plot shows that the sample mean of CCDF agrees with the benhcmark. This is consistent with common findings and not surprising because SS is asymptotically unbiased for sufficiently large $N$. 
%
%----------------------------------------------
% other rows: sensitivity
%----------------------------------------------
The new investigation that matters lies in the rows below. 
For the plots in the middle column, all lines are visually close, i.e., the sample mean of SS and its $\pm 1 \sigma$ bounds all gather around the benchmark. This is largely attributed to the fact that the sensitivity in the middle column is dominated by the PDF of $Y$. The PDF estimate derives it quality from the CCDF estimate, which is of good quality as evidenced from the top-left plot. 
%--------------- 
% right plot
%--------------- 
A more critical assessment lies in the plots of the right column, the sensitivity measure recommended in this work. 
For $y$ beyond $y_2$, i.e., the middle vertical green line, a visual departure of the blue dots (SS mean) from the red line (benchmark) can be seen to increase with $y$, implying an increasing bias. 
The gap between the $\pm 1\sigma$ bounds also widens, not suprising because it runs out of samples for $y>y_3$ (the rightmost vertical line). 
For $y< y_3$, the growth of variance is somewhat suppressed, by virtue of SS that populates samples up to around $y_3$.  
The bias can be further reduced by using a smaller kernel width, though at the expense of widening the $\pm 1 \sigma$ gap.
The sensitivity estimate at the left end (small $y$) of the plot diverges because it runs out of samples there. It is immaterial because that region of low thresholds is not of concern in applications.

%==============================================
%----------------------------------------------------------
% eg01 - stats
%----------------------------------------------------------
%==============================================
%##################
\FloatBarrier
%##################

%==============================
\subsection{Buckling load of shear building}\label{sec:buckle}
%==============================
Consider the buckling of a structure with linear-elastic stiffness matrix $\bK$, and geomatric stiffness matrix $\bK_g$ that accounts for first order P-Delta effect. 
To assess buckling risk, we define the `response' as
\begin{align}\label{eq:buckle_Y}
Y = \frac{\lambda_0}{\lambda}
\end{align}
where $\lambda_0$ is a normalizing constant, see Section~\ref{sec:buckle_results} later for specific value; $\lambda$ is the critical buckling load factor (the higher the better), equal to the smallest eigenvalue of the generalized eigenvalue problem:
\begin{align}\label{eq:buckle_eig}
\bK \bu = \lambda\, \bK_g \bu
\end{align} 
where $\bu$ is the eigenvector, i.e., buckling shape. 
For an $n$-storied shear building with lateral story stiffness $k_i$ ($1\le i\le n_s$),
\begin{align}\label{eq:buckle_K}
\bK = 
\begin{bmatrix}
k_1 + k_2 & -k_2 & & &\\
-k_2 & k_2 + k_3 & -k_3 & &\\
 & -k_3 & \ddots & \ddots &\\
  & & \ddots & k_{n-1}+k_n & -k_n\\
  & & & -k_n & k_n
\end{bmatrix}
\end{align}
The geomatrix stiffness matrix $\bK_g$ has the same form as $\bK$ except that $k_i$ is replaced by $W_i/H_i$, where $W_i$ is the floor load and $H_i$ is the story height.

%=========================
\subsubsection{Response derivatives}
%=========================
%The response gradient $G = \partial Y/\partial \hp$ in \eqref{eq:G} required for the theory is related to that of $\lambda$. 
Differentiating \eqref{eq:buckle_Y} gives, for a generic sensitivity parameter $\hp$, 
\begin{align}\label{eq:buckle_dY}
\frac{\partial Y}{\partial \hp} = - \frac{\lambda_0}{\lambda^2} \frac{\partial \lambda}{\partial \hp}
\end{align}
For general $\bK$ and $\bK_g$, the derivative of eigenvalue $\lambda$ and its eigenvector $\bu$ (though not required here) can be obtained by solving a matrix equation of augmented dimension~\cite{Lee1997}:
\begin{align}\label{eq:buckle_deriv}
\begin{bmatrix}
\bK - \lambda \bK_g & - \bK_g \bu\\
-\bu^T \bK_g & 0
\end{bmatrix}
\begin{bmatrix}
\partial \bu /\partial \hp \\
\partial \lambda /\partial \hp
\end{bmatrix}
=
\begin{bmatrix}
-\left( \partial \bK /\partial \hp - \lambda\, \partial \bK_g /\partial \hp \right) \bu\\
\frac{1}{2}\bu^T (\partial \bK_g /\partial \hp) \bu
\end{bmatrix}
\end{align}
See a review on eigenvalue problem sensitivity~\cite{Lin2020}.
In \eqref{eq:buckle_deriv}, the upper partition results from taking derivative of the eigenvalue equation in \eqref{eq:buckle_eig}. 
The lower partition results from taking derivative of the scaling constraint $\bu^T \bK_g \bu = $ constant. 
For non-singular $\bK$ and $\bK_g$ it can be shown that the coefficient matrix on the LHS is always invertible, and hence the stability of numerical solution is guaranteed; see Section 3.2 of \cite{Lee1997}. 
Also, since the coefficient matrix is symmetric, efficient algorithms such as LU factorization can be used for solution. 
For a given eigen pair $(\lambda,\bu)$, the coefficient matrix remains the same regardless of the sensitivity parameter $\hp$ under question, and so the factorization, which can be time-consuming for large structures, only needs to be performed once.
%
%This sensivitiy technique was applied to improving efficiency in Bayesian structural model updating~\cite{Zhu2023}. 
%

%=========================
\subsubsection{Uncertainty modeling}
%=========================
Let the floor loads $W_i$ be independent Lognormal with mean $w_i$ and coefficient of variation (c.o.v.\@ = standard deviation/mean) $\delta_i$. To fit into the context of this work where the PDF of input random variables does not depend on sensitivity parameters, we represent the $W_i$'s in terms of i.i.d.\@ $N(0,1)$ variates $X_i$'s as
\begin{align}\label{eq:buckle_Wi}
W_i = w_i \, e^{a_i + b_i X_i}
\end{align}
where
\begin{align}\label{eq:root}
a_i = - \ln \sqrt{1+\delta_i^2}
\qquad
b_i = \sqrt{\ln(1+\delta_i^2)}
\end{align}
are the mean and standard deviation of the $\ln(\cdot)$ of a Lognormal variate with mean $1$ and standard deviation $\delta_i$.
% 
%They can be determined using the relationshp between the mean and standard deviation ($\mu,\sigma$) of a Lognormal variate and those ($u,s$) of its natural logarithm, i.e., 
%{\color{red}
%
%\begin{align}\label{eq:root}
%u = \ln\left( \frac{\mu^2}{\sqrt{\mu^2 + \sigma^2}}\right)
%\qquad
%s^2 = \ln\left( 1 + \frac{\sigma^2}{\mu^2}\right)
%\end{align}
%%
%\begin{align}
%\mu = e^{a+b^2/2}
%\qquad
%\sigma^2 = e^{2a+b^2} (e^{b^2}-1 )
%\end{align}
%%
%}
%
%Note that in our case $\mu = 1$ and $\sigma = \delta_i$.

%======================
\subsubsection{Analytical solution for benchmarking}\label{sec:buckle_anal}
%======================
For general structures, $\lambda$ requires solving the eigenvalue problem in \eqref{eq:buckle_eig}. 
For a shear building with $\bK$ and $\bK_g$ of the form in \eqref{eq:buckle_K}, however, all eigenvalues $\lambda_i$ and eigenvectors $\bu_i$ can be obtained analytically as 
\begin{align}\label{eq:buckle_lambdai}
\lambda_i = \frac{k_i H_i}{W_i}
\end{align}
and $\bu_i$ is a vector equal to 1 at all entries from $i$ to $n$, and zero elsewhere. Essentially, the $i$th mode corresponds to buckling in the $i$th story only. 
This analytical result can be verified by noting that $\bK \bu_i$ gives a vector equal to $k_i$ at the $i$th entry and zero elsewhere; the same for $\bK_g\bu_i$ except that the $i$th entry is $W_i/H_i$. The two vectors are therefore colinear, implying that $\bu_i$ is an eigenvector. The $i$th row of $\bK \bu_i = \lambda_i \bK_g \bu_i$ reads $k_i = \lambda_i W_i/H_i$, which gives \eqref{eq:buckle_lambdai}.

By noting $\lambda = \min_{1\le i \le n} \lambda_i$ and using \eqref{eq:buckle_Wi}, $Y$ in \eqref{eq:buckle_Y} is now given by
\begin{align}\label{eq:buckle_Y_shear_general}
Y = \lambda_0 \max_{1\le i \le n} \left\{ \frac{w_i}{k_i H_i} e^{a_i+b_i X_i}\right\}
\end{align} 
Since $\{Y<y\}$ is equivalent to all terms in the $\max\{\cdot\}$ less than $y$, and the terms are independent, we have
\begin{align}\label{eq:buckle_F}
P(Y<y) &= \prod_{i=1}^n \Phi(\bar{x}_i)
\end{align}
where
\begin{align}
\bar{x}_i = \frac{1}{b_i}\left[ \ln\left( \frac{k_i H_i y}{w_i \lambda_0}\right) - a_i\right]
\end{align}
Using \eqref{eq:buckle_F}, $P(Y\ge y) = 1-P(Y<y)$ and its sensitivities can be obtained analytically.
See \eqref{eq:buckle_F_num}, \eqref{eq:buckle_dF_num01} and \eqref{eq:buckle_dF_num02} later in the numerical example.
%%
%{\color{red}
%It should be note that $\partial Y/\partial \hp$ exists and continuous except at the points where the critical mode switch from one story to another, which has probability zero. This is admissible in the theory?}

%==============================
\subsubsection{Illustrative scenario}\label{sec:buckle_results}
%==============================
For numerical investigation, consider a shear building with $n=5$ stories and the following nominal properties: uniform story height $H = 3.5$~m and stiffness $k = 250$~kN/mm, and i.i.d.\@ Lognormal floor loads $W_i$ with uniform mean $w=100$~kN and c.o.v.\@ 10\%. 
The normalizing constant $\lambda_0$ in \eqref{eq:buckle_Y} is taken as the value of $\lambda$ when $W_i$'s are at their mean value. Equation \eqref{eq:buckle_lambdai} gives $\lambda_0 = 250\cdot 3500/1000 = 875$.

We will investigate the sensitivity of $P(Y\ge y)$ w.r.t.\@ the common mean of floor load $w \equiv \hp_1$ and the second story stiffness $k_2 \equiv \hp_2$, both about their nominal values (100~kN, 250~kN/mm). All other quantities are fixed at their nominal values. 
The different nature of $\hp_1$ and $\hp_2$ serves to demonstrate applicability to parameters that affect the distribution of input random variables ($\hp_1$) and those that affect the response directly ($\hp_2$).

%-----------------------------
% use SS and theory
%-----------------------------
%The failure probabilities and sensitivities will be calculated using SS and the proposed methodology in \eqref{eq:J_ss}, and then compare them with the analytical solutions in \ref{sec:buckle_anal}.
%

When applying SS for sensitivity, to be consistent wtih typical applications, $\lambda$ will be obtained by solving the eigenvalue problem in \eqref{eq:buckle_eig}, and its derivative $\partial \lambda/\partial \hp$ by solving \eqref{eq:buckle_deriv}.
In the latter equation, it can be reasoned from \eqref{eq:buckle_K} that $\partial \bK/\partial \hp_2$ has the same form as $\bK$ with $k_2$ replaced by 1 and all other $k_i$'s by 0. 
On the other hand, recall that $\bK_g$ has the same form as $\bK$ except that $k_i$ is replaced by $W_i/H_i$, and $W_i = \hp_1 \exp(a + b X_i)$. 
This implies that $\partial \bK_g/\partial \hp_1$ has the same form as $\bK$ except that $k_i$ is replaced by $H^{-1}\exp(a + b X_i)$.
%

%--------------------
% analytical
%--------------------
For benchmarking, the analytical solution in \eqref{eq:buckle_Y_shear_general} is given by ($n=5$)
\begin{align}\label{eq:buckle_Y_shear02}
Y = \lambda_0 \max \left\{ \frac{\hp_1}{k H} e^{a+b X_i}, \frac{\hp_1}{\hp_2 H} e^{a+b X_2},\cdots, \frac{\hp_1}{k H} e^{a+b X_n}\right\}
\end{align} 
where $a = -4.975\times 10^{-3}$ and $b = 0.9975\times 10^{-3}$ (4 digits) are obtained from \eqref{eq:root} with mean 1 and c.o.v.\@ 10\%.
Equation \eqref{eq:buckle_F} then gives
{
\begin{align}\label{eq:buckle_F_num}
P(Y < y) =  \Phi(\bar{x}_1)^{n-1} \Phi(\bar{x}_2)
\end{align}
}
where
\begin{align}
\bar{x}_1 &= \frac{1}{b}\left[ \ln\left( \frac{k H y}{\hp_1 \lambda_0}\right) - a\right]\\
\bar{x}_2 &= \frac{1}{b}\left[ \ln\left( \frac{\hp_2 H y}{\hp_1 \lambda_0}\right) - a\right]
\end{align}
%
%Note that $\hp_1$ affects $\bar{x}_1$ only, but $\hp_2$ affects all $\bar{x}_i$.
%
Taking derivatives gives the sensitivities
{
\begin{align}
%-----------------------------------------------
\frac{\partial P(Y< y)}{\partial \hp_1} &=
- \frac{1}{b\hp_1} \left[(n-1)\frac{\phi(\bar{x}_1)}{\Phi(\bar{x}_1)} + \frac{\phi(\bar{x}_2)}{\Phi(\bar{x}_2)}\right]  P(Y<y)
\label{eq:buckle_dF_num01}\\
%-----------------------------------------------
\frac{\partial P(Y< y)}{\partial \hp_2} &=
\frac{1}{b\hp_2}  \frac{\phi(\bar{x}_2)}{\Phi(\bar{x}_2)} P(Y<y) 
\label{eq:buckle_dF_num02}
%-----------------------------------------------
\end{align}
}
Clearly, $P(Y\ge y) = 1-P(Y<y)$ and $\partial P(Y\ge y)/\partial \hp_i = - \partial P(Y< y)/\partial \hp_i$.

%------------------------------------------------------------
\subsubsection{Numerical results}
\begin{figure}[h]
{
\includegraphics[width=\textwidth]{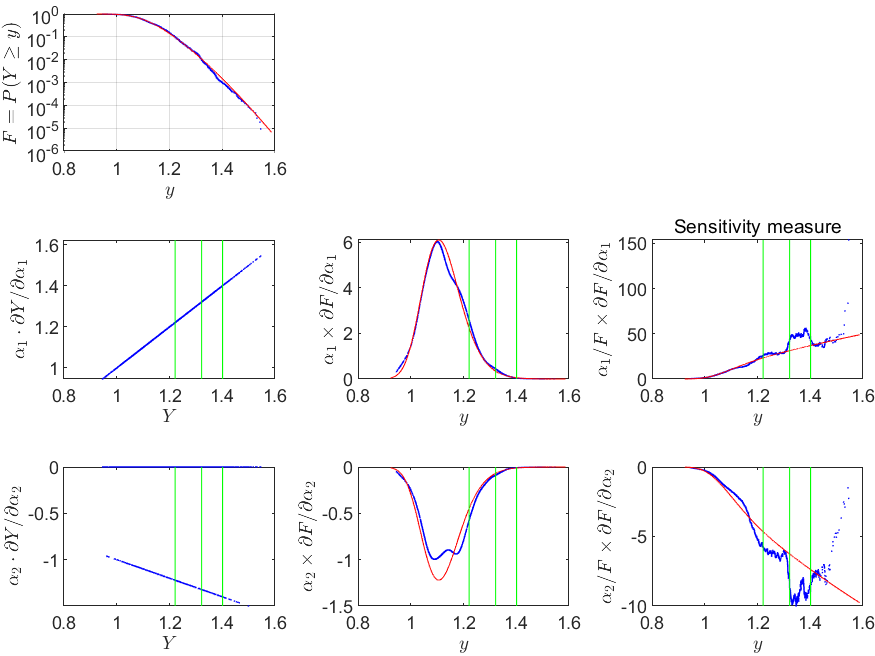}
\caption{Example 2 (Shear building buckling). Results of a typical run. Same legend as Figure~\ref{fig:eg01_single}}
\label{fig:buckle_single}
}
\end{figure}

\begin{figure}[h]
{
\includegraphics[width=\textwidth]{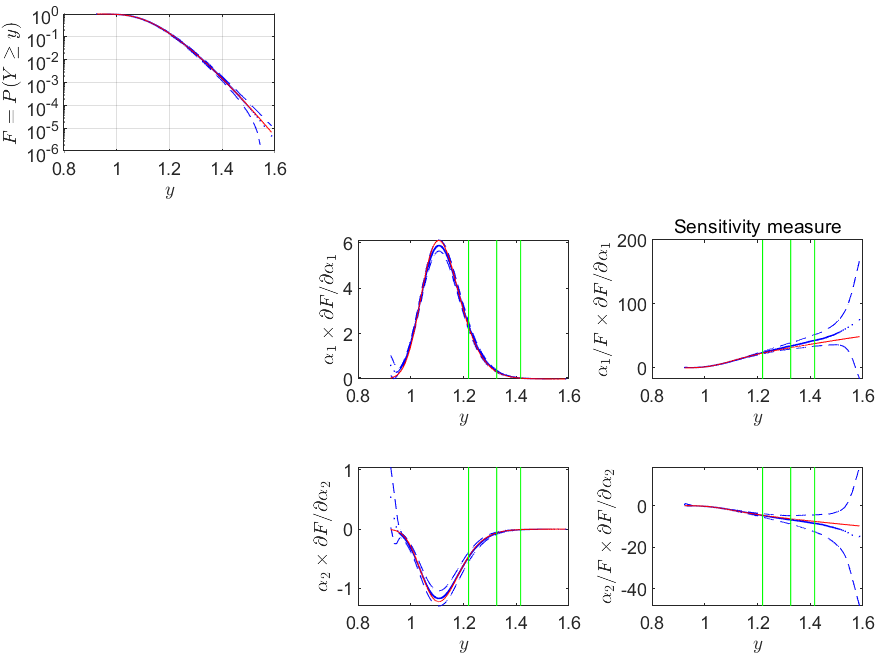}
\caption{Example 2 (Shear building buckling). Sample mean (blue solid line) and $\pm 1\sigma$ bounds (blue dashed line) from 1000 SS runs. Same legend as Figure~\ref{fig:eg01_stats}}
\label{fig:buckle_stats}
}
\end{figure}

Figure~\ref{fig:buckle_single} shows the results of a typical SS run, in the same manner as Figure~\ref{fig:eg01_single}.  
The CCDF in the top-left plot agrees with benchmark. 
%---------------------------------
% alpha 1, left
%---------------------------------
In the second row, the left plot shows that the response gradient of $\hp_1$ is fully correlated with the response. This can be seen by taking $\hp_1$ outside the $\max\{\cdot\}$ in \eqref{eq:buckle_Y_shear02} so that it reads $Y = \hp_1 \times (\cdot)$ and hence $\partial Y/\partial \hp_1 = (\cdot)$ differs from $Y$ only by a deterministic factor ($\hp_1$). 
%---------------------------------
% alpha 1, middle & right
%---------------------------------
The quality of sensitivity estimate for $\hp_1$ in the middle and right plot is similar to that for $\hp_1$ or $\hp_2$ in Figure~\ref{fig:eg01_single} for Example~1. 
%---------------------------------
% alpha 2, middle
%---------------------------------
The sensitivity estimate for $\hp_2$ has a lower quality, as evidence from the middle and right plot at the bottom.
This is attributed to the fact that the response gradient has a more complicated structure (hence estimation variance), as seen in the bottom left plot. 
Although the points at the lower part of the plot show a full correlation similar to what was seen for $\hp_1$, there are many points at zero response gradient. 
This can be explained by \eqref{eq:buckle_Y_shear02}. When the second term is greater than all other terms, $Y = \hp_2^{-1}\times (\cdot)$ and so $\partial Y/\partial \hp_2 = -\hp_2^{-2}\times (\cdot)$ is fully negatively correlated with $Y$. Otherwise $Y$ does not depend on $\hp_2$ and so $\partial Y/\partial \hp_2 = 0$.
Intuitively, the second story stiffness ($\hp_2$) matters to the critical buckling load only when the second story is most critical. 
In the present case where $\hp_1=\hp_2=1$, the terms in the $\max\{\cdot\}$ are i.i.d.\@ and so the chance of the second story being critical is $1/n_s = 1/5$. In other words there is a $80\%$ chance that $\partial Y/\partial \hp_2 = 0$. This creates a disparity in the samples of response gradient, inflating the estimation variance in a similar manner as averaging indicator function values of rare events.   
Figure~\ref{fig:buckle_stats} shows the sample mean and $\pm 1\sigma $ bounds of sensitivity estimates, in the same manner as Figure~\ref{fig:eg01_stats}.
The statistics are consistent with the single run results in Figure~\ref{fig:buckle_single}.
%
%Judging difficulty by the $\pm 1\sigma $ bounds, $\hp_1$ in this example is similar to the $\hp_i$'s in Example~1; and $\hp_2$ is more difficult. 

%==============================================
%----------------------------------------------------------
% buckle - single run
%----------------------------------------------------------
%==============================================
%##################
\FloatBarrier
%##################

%==============================================
%----------------------------------------------------------
% buckle - stats
%----------------------------------------------------------
%==============================================
%##################
\FloatBarrier
%##################

%==============================
\subsection{Linear SDOF first passage problem}\label{sec:sdof}
%==============================
Consider a linear SDOF structure subjected to white noise excitation. The response $Y$ is defined as 
\begin{align}\label{eq:sdof_Y}
Y = \max_{0\le t\le T} |u(t)|
\end{align}
where $T = 20$~s is the duration; $u(t)$ is the displacement starting from rest $u(0) = \dot{u}(0) = 0$ and thereafter follows the equation of motion
\begin{align}\label{eq:sdof_EoM}
\ddot{u}(t) + 2\zeta\omega \dot{u}(t) + \omega^2 u(t) = W(t)
\end{align}
with $\omega = 2\pi $ (rad/s, natural frequency), $\zeta = $~1\% (damping ratio), and $W(t)$ is white noise with two-sided power spectral density (PSD) $S = 0.86 N^2/(rad/s)$.
This example was considered recently in \cite{Au2025} when investigating the optimal choice of correlation parameter in conditional sampling MCMC, and in \cite{Au2026_dR1} on a theory for optimal MCMC for rare events. 
Here, the sensitivty of $P(Y\ge y)$ w.r.t.\@ $\zeta$ and $\omega$ will be investigated. 
From \eqref{eq:sdof_Y}, for a generic $\hp$,
\begin{align}\label{eq:sdof_dY}
\frac{\partial Y}{\partial \hp} = \chi\, \frac{\partial u(t_*)}{\partial \hp}
\end{align}
where $t_*$ is the peak response time, i.e., when $|u|$ is maximum, and $\chi$ is the sign of $u(t_*)$ that accounts for the effect of $|\cdot|$. 
For general $t$, $\partial u(t)/\partial \hp$ can be obtained by solving an equation resulting from differentiating \eqref{eq:sdof_EoM} w.r.t.\@ $\hp$. 
Let $u_{\zeta}(t) = \partial u(t)/\partial \zeta$ and $u_{\omega}(t) = \partial u(t)/\partial \omega$. Then (omitting $t$)
\begin{gather}
\ddot{u}_{\zeta} + 2\zeta \omega \dot{u}_{\zeta} + \omega^2 u_{\zeta} = - 2\omega \dot{u}
\label{eq:sdof_du01}\\
\ddot{u}_{\omega} + 2\zeta \omega \dot{u}_{\omega} + \omega^2 u_{\omega} = - 2\zeta \dot{u} - 2\omega u
\label{eq:sdof_du02}
\end{gather}
These equations differ by the RHS only, and their LHS have the same functional form as the LHS of \eqref{eq:sdof_EoM}.
When $u(t)$ is given, their RHS are known and they can be solved in the same way as \eqref{eq:sdof_EoM}. 
Computation is performed in discrete time at a time interval of $\Delta t= 0.05$~s and using {\tt lsim} in Matlab. 
There are $n = T/\Delta t= 400$ input random variables in the discrete white noise $W_i = \sqrt{2\pi S/\Delta t}\, X_{i+1}$ ($0\le i\le n-1$), where $\{X_i\}_{i=1}^n$ are the i.i.d.\@ N(0,1) input random variables.

%------------------------------------------------------------
\subsubsection{Numerical investigation}
%------------------------------------------------------------
%Similar to prevous examples, SS runs are performed for $m = 3$ levels, $p_0 = 0.1$ and $N = 1000$ samples at each level. 
%
In this example, analytical solution is not available. 
The benchmark for the CCDF of $Y$ is obtained by direct Monte Carlo with $10^7$ i.i.d.\@ samples. Using the same set of input random variables $\bX$, i.e., common random numbers, the sensitivity benchmark is obtained by central difference with a relative step of 1\%. This requires an additoinal $10^7$ response function evaluations for the forward step, and another $10^7$ for the backward step. 

\begin{figure}[h]
{
\includegraphics[width=\textwidth]{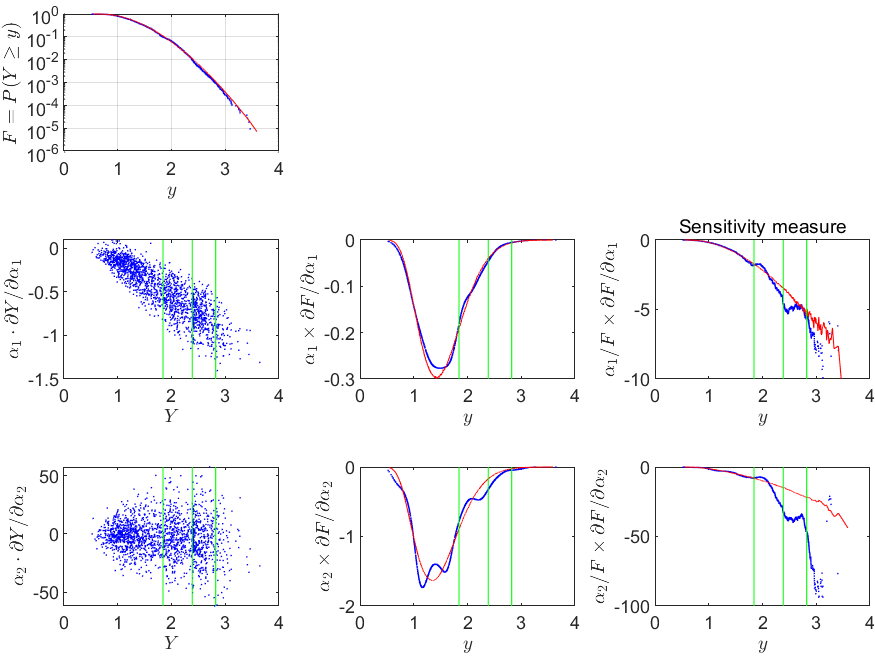}
\caption{Example 3 (SDOF first passage). Results of a typical run. Same legend as Figure~\ref{fig:eg01_single}}
\label{fig:sdof_single}
}
\end{figure}

Figure~\ref{fig:sdof_single} shows the results for a typical run, in the same way as Figure~\ref{fig:eg01_single}. 
In the top-left plot the CCDF estimate by SS agrees with the benchmark, as expected. 
%----------------------------------------
% alpha 1: damping ratio
%----------------------------------------
In the middle row, the left plot shows that the response gradient of $\hp_1$ (damping ratio) is negative, and is negatively correlated with the response. The former is intuitive; the latter is not trivial. 
The quality of sensitivity estimate in the middle and right plot is similar to those in Example~1, suggesting similar difficulty. 
%----------------------------------------
% alpha 2: natural frequency
%----------------------------------------
This is not the case for $\hp_2$ (natural frequency) in the bottom row. 
The bottom middle plot shows a spurious fluctuation with $y$; the bottom right plot shows a significant departure from the benchmark as $y$ increases. 
The difficulty with $\hp_2$ this time is more subtle than $\hp_2$ in Example 2, and the mechanism is not well-understood at the time of writing. 
Visual inspection of the scatter plot in the bottom left would suggest that the response gradient may be uncorrelated from the response. 
The bottom right plot defies this, as the sensitivity there has a similar order of magntiude as $\hp_1$ (damping ratio). 
In this sense, the natural frequency ($\hp_2$) is a more difficult parameter than the damping ratio ($\hp_1$). 
%
%The difficulty is more subtle than $\hp_2$ in Example~2, with apparently low correlation between the response and its gradient but still carrying significant sensitivity on failure probability. 

\begin{figure}[h]
{
\includegraphics[width=\textwidth]{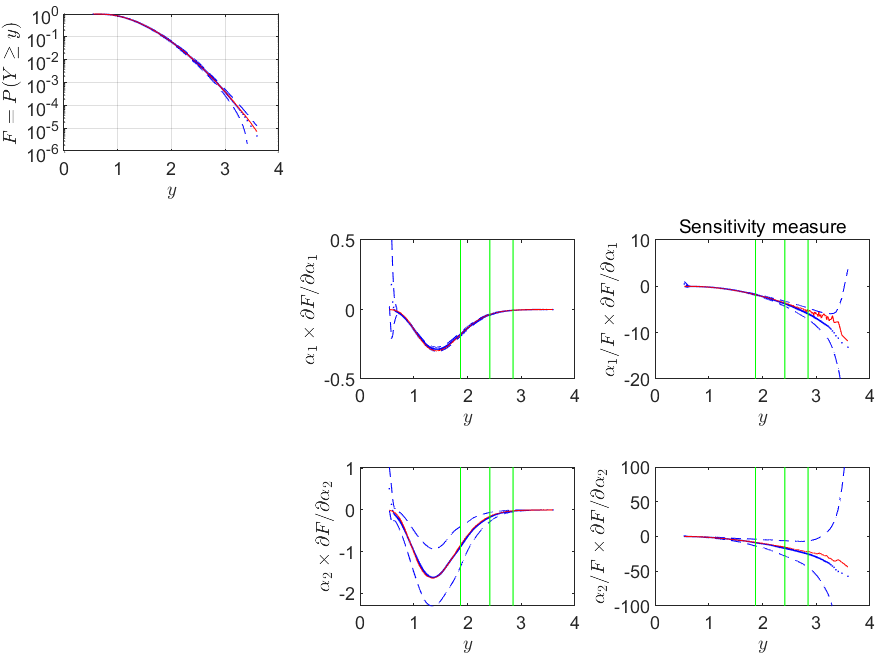}
\caption{Example 3 (SDOF first passage). Sample mean (blue solid line) and $\pm 1 \sigma $ bounds (blue dashed line) from 1000 SS runs. Same legend as Figure~\ref{fig:eg01_stats}}
\label{fig:sdof_stats}
}
\end{figure}

Figure~\ref{fig:sdof_stats} shows the sample mean and $\pm 1 \sigma$ bounds of sensitivity estimates, in the same way as Figure~\ref{fig:eg01_stats}. The results are consistent with those in Figure~\ref{fig:sdof_single} for a single run. 
The $\pm 1 \sigma$ gap for the sensitivity of $\hp_2$ (natural frequency) is one order of magnitude wider than that for $\hp_1$ (damping ratio). 

%------------------------------------------------------------

%==============================================
%----------------------------------------------------------
% sdof - single run
%----------------------------------------------------------
%==============================================
%##################
\FloatBarrier
%##################

%==============================================
%----------------------------------------------------------
% sdof - stats
%----------------------------------------------------------
%==============================================
%##################
\FloatBarrier
%##################

%==============================
\subsection{Pile serviceability design}\label{sec:pile}
%==============================
This example is adapted from Section~6 of \cite{Wang2013} that investigated reliability-based design with SS implemented in spreadsheets. 
Consider assessing the `failure' as inadequacy of a concrete pile in loose sand against serviceability limit state (SLS) of excessive settlement. 
The pile has a circular section with diameter $B = 0.9$~m and depth $D = 8$~m into the soil. 
The latter has uncertainty in the friction angle $\phi'$ that depends on depth $z$. 
The dash here indicates that the quantity is associated with effective stress, a convention in geotechnical engineering.
The friction angle is modeled as a spatially stationary Lognormal random field with mean $\mu$ and c.o.v.\@ $\delta$. The correlation between $\ln \phi_i$ at depth $z_i$ and $\ln \phi_j$ at $z_j$ is assumed to be
\begin{align}\label{eq:Rij}
R_{ij} = \exp\left(-\frac{2}{\lambda} |z_i-z_j|\right)
\end{align}  
where $\lambda = $~4~m is a correlation length. 
For computation, a soil depth of $12$~m is considered, which is sufficient to cover the relevant influence zone of resisance (see later). Dividing it into differential segments of equal thickness $d = 0.1$~m, the uncertainty in frictional angle is characterized by the values $\{\phi'_i\}_{i=1}^n$ at depths $z_i = (i-1/2)d$, where $n = 120$ is the number of random variables. 
The random vector $\bphi' = [\phi'_1,\dots,\phi'_n]^T$ can be represented by a i.i.d.\@ N(0,1) vector $\bX$ by
\begin{align}\label{eq:phi}
\bphi' = \mu \exp\left( u \ones + s \bL \bX \right)
\end{align} 
where $\exp(\cdot)$ operates entry by entry; $\ones$ denotes an $n\times 1$ vector of ones; 
$u = - \ln \sqrt{1+\delta^2}$ and $s = \sqrt{\ln(1+\delta^2)}$ are respectively the mean and standard deviation of the $\ln(\cdot)$ of a Lognormal variate with mean 1 and standard deviation $\delta$; and $\bL$ is the lower triangular matrix from Cholesky decomposition $\bR = \bL \bL^T$, where $\bR$ is the $n\times n$ correlation matrix of $\{\ln\phi'_i\}_{i=1}^n$ with the $(i,j)$-entry given by \eqref{eq:Rij}.

The response $Y$ is defined as the ratio of load to effective resistance, the higher the worse. 
We will investigate the sensitivity of $P(Y\ge y)$ w.r.t.\@ the pile diameter and the mean of the friction angle, i.e., $\hp_1 \equiv B$ and $\hp_2 \equiv \mu$. 
Following the notation in Section~6 of \cite{Wang2013}, the response is given by
\begin{align}
Y = \frac{F_{50}}{Q_{sls}}
\end{align}  
where $F_{50}$ is the design load,   
\begin{gather}
\label{eq:Qsls}
Q_{sls} = 0.625\, a \left(\frac{y_a}{B}\right)^b \left(Q_{side} + Q_{tip} - W\right)\\
%------------------------------------------------------------------------------------------
\label{eq:Qside}
Q_{side} = \pi B\, d\, (K/K_0)_n K_0 \sum_{i=1}^{N_s} \sigma_{i}' \tan \phi_i'\\
%------------------------------------------------------------------------------------------
\label{eq:Qtip}
Q_{tip} = 0.25\pi B^2 \left[ 0.5 B(\gamma-\gamma_w) N_{\gamma} \zeta_{\gamma s}\zeta_{\gamma d} \zeta_{\gamma r} + D (\gamma - \gamma_w) N_q \zeta_{qs} \zeta_{qd} \zeta_{qr}\right]\\
%------------------------------------------------------------------------------------------
\label{eq:W}
W = 0.25\pi B^2 D\, (\gamma_{c} - \gamma_w)
%------------------------------------------------------------------------------------------
\end{gather}
are respectively the resistance capacity at SLS, side resistance, tip resistance, and effective pile weight.  
%----------------------------------
% different categories
%----------------------------------
The parameters appearing in these equations are summarized in Table~\ref{tab:pile}, categorized by their nature. 
The top category shows the source of uncertainty, i.e., $\{\phi'_i\}_{i=1}^n$. 
The next category contains the ones affected by $\{\phi'_i\}_{i=1}^n$, essentially through the mean friction angle $\phibar'$. 
It also contains the pile diameter $B$, which is a sensitivity parameter ($\hp_1$). 
The parameters at the bottom category are constants.

%=====================
% Table: pile example
%=====================
\begin{table}[h]
\caption{Parameters in \eqref{eq:Qsls} to \eqref{eq:W} of the pile example in Section~\ref{sec:pile}}
\label{tab:pile}
\begin{center}\scriptsize
\begin{tabular}{m{7em} m{18em} m{18em}}
%---------------------------------------------------------
\hline
Symbol & Value/formula & Description\\
\hline
%---------------------------------------------------------
$\phi'_i$, $1\le i\le n$ & Lognormal random, mean $\mu = $~0.5585~rad (32$^\circ$), c.o.v.\@ $\delta = $~17\%, correlation length $\lambda =$ 4~m. See \eqref{eq:phi}  & Friction angle at depth $z_i = (i-1/2)\, d$; $n = 120$ (no. of discretized layers), $d = 0.1$~m (layer thickness)\\
\hline
%---------------------------------------------------------
$\phibar'$ &  & 
Average $\phi$ in tip resistance influence zone, from $\min\{8B,D\}$ above to $3.5B$ below pile tip\\
$B$ & 0.9~m & Pile diameter\\
$N_q$ & $\tan^2(\pi/4 + \phibar'/2) \exp(\pi \tan \phibar')$ & Bearing capacity factor (load)\\
$N_{\gamma}$ & $2 (N_q +1 ) \tan \phibar'$ & Bearing capacity factor (weight)\\
$\zeta_{qs}$ & $1+\tan\phibar'$ & Correction factor\\
$\zeta_{qd}$ & $1+2(\tan\phibar')(1-\sin \phibar')^2 \tan^{-1}(D/B)$ & Correction factor\\
\hline
%---------------------------------------------------------
$N_s$ & Integer part of $D/d$ & No. of differential layers in side resistance influence zone of length $D$\\
$\sigma_i'$ & $(\gamma-\gamma_w) z_i$ & Effective stress at depth $z_i$\\ 
$F_{50}$ & 800~kN & Design load\\
$D$ & 8~m & Pile depth\\
$y_a$ & 25~mm & Allowable displacement\\
$\gamma$ & 20~kN/m$^3$ & Unit weight of soil\\
$\gamma_w$ & 9.81~kN/m$^3$ & Unit weight of water\\
$\gamma_{c}$ & 24~kN/m$^3$ & Unit weight of concrete\\
$(K/K_0)_n$ & 1 & Nominal operative in-situ horizontal stress coefficient ratio\\
$K_0$ & 1 & At-rest coefficient of horizontal soil stress\\
$a,b$ & 4, 0.4 & Constants\\
$\zeta_{\gamma s},\zeta_{\gamma d},\zeta_{\gamma r},\zeta_{qr}$ & 0.6, 1, 1, 1 & Correction factors\\
\hline
%---------------------------------------------------------
\end{tabular}
\end{center}
\end{table}

%##################
\FloatBarrier
%##################

%--------------------------------------------------
% response gradient by finite difference
%--------------------------------------------------
In this example, the response derivatives $\partial Y/\partial \hp_i$ are computed by central difference with a relative step size of 0.1\%. This is the case in applications where only response function values are available. 
Strictly speaking, this will induce an additional bias in the resulting sensitivity estimate, although it is negligible compared to the bias due to kernel smoothing. 
%----------------------------------------------
% how B affect the Q's
%----------------------------------------------
Note that $B$ ($\equiv\hp_1$) affects $Q_{sls}$, $Q_{side}$ and $Q_{tip}$ directly through its appearance in \eqref{eq:Qsls} to \eqref{eq:Qtip}, and indirectly through $\phibar'$, because the latter is averaged over the tip resistance influence zone from $\min\{8B,D\}$ above to $3.5B$ below the pile tip; see Table~\ref{tab:pile}. 
%----------------------------------------------
% how mu affect the Q's
%----------------------------------------------
On the other hand, $\mu$ ($\equiv\hp_2$) is the mean value of $\{\phi'_i\}_{i=1}^{n}$, which in turn affects $Q_{side}$ through $\tan(\cdot)$ in the sum of \eqref{eq:Qside}, and $Q_{tip}$ in \eqref{eq:Qtip} through $\phibar'$ that affects $N_q$, $N_{\gamma}$, $\zeta_{qs}$ and $\zeta_{qd}$, also in a trigonometric manner; see the second category in Table~\ref{tab:pile}.

%------------------------------------------------------------
\subsubsection{Numerical results}
\begin{figure}[h]
{
\includegraphics[width=\textwidth]{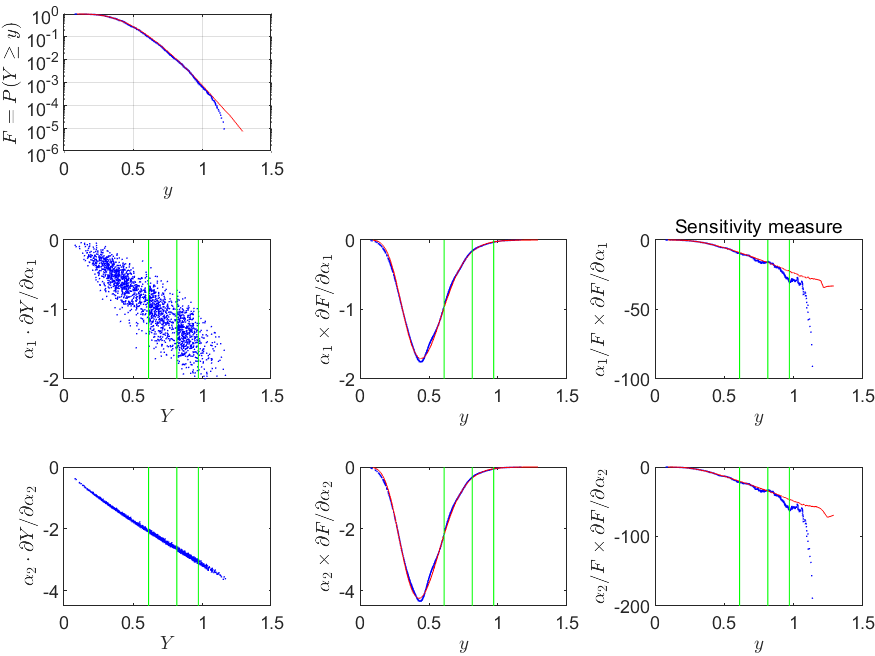}
\caption{Example 4 (pile design). Results of a typical run. Same legend as Figure~\ref{fig:eg01_single}}
\label{fig:pile_single}
}
\end{figure}

\begin{figure}[h]
{
\includegraphics[width=\textwidth]{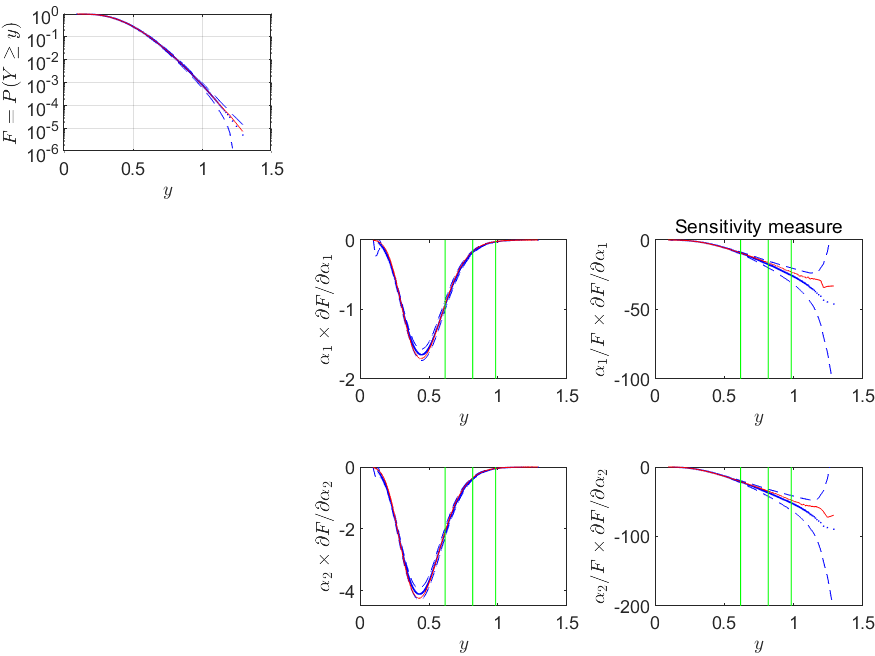}
\caption{Example 4 (pile design). Sample mean (blue solid line) and $\pm 1 \sigma $ bounds (blue dashed line) from 1000 SS runs. Same legend as Figure~\ref{fig:eg01_stats}}
\label{fig:pile_stats}
}
\end{figure}

%------------------------------------------------------------
Figures~\ref{fig:pile_single} and \ref{fig:pile_stats} show the results of a single run and statistics from 1000 indepenent runs, analogous to Figures~\ref{fig:eg01_single} and \ref{fig:eg01_stats} of Example~1. 
%-------------------------
% benchmark
%-------------------------
Similar to Example~3 (SDOF first passage), analytical solution for failure probability or sensitivity is not available. The benchmark is taken as the central difference (relative step size 1\%) of Monte Carlo estimates for $P(Y\ge y)$, each based on $10^7$ samples. 
Generally, the quality of sensitivity estimates is similar to that of Example~1, reflecting sensitivity parameters of similar difficulty. 
For both $\hp_1$ ($\equiv B$) and $\hp_2$ ($\equiv \mu$), the response is negatively correlated with its gradient. The former has a larger scatter. These are not trivial to explain. 
Both parameters have a reliability sensitivity in the order of a few tens. 
At a failure probability of 0.1\%, i.e., $y\approx 1$, the sensitivities (right column) of $\hp_1$ and $\hp_2$ are about 25 and 50, respectively. This implies that, e.g., a 1\% change in $\hp_1$ will lead to about 25\% change in $P(Y\ge y)$. Also, $\hp_2$ is twice as sensitive than $\hp_1$. 

%==============================================
%----------------------------------------------------------
% pile - single run
%----------------------------------------------------------
%==============================================
%##################
\FloatBarrier
%##################

%==============================================
%----------------------------------------------------------
% sdof - stats
%----------------------------------------------------------
%==============================================
%##################
\FloatBarrier
%##################

%===============================================
\section{Conclusions}\label{sec:conclusions}
%===============================================
As the key theoretical contribution, this work has derived a formula that expresses reliability sensitivity $\partial P(Y\ge y)/\partial \hp$ in terms of response gradient $\partial Y/\partial \hp$. See \eqref{eq:dF_intg} as an integral and \eqref{eq:dF_E} as a conditional expectation. 
Based on the formula a Monte Carlo method has been developed to calculate $\partial P(Y\ge y)/\partial \hp$ together with $P(Y\ge y)$ in the same run and for all generated values of $y$. See \eqref{eq:dF_SS}, which leverages Subset Simulation for generating rare event samples, and kernel smoothing for resolving the issue of zero-probability conditioning. 
The latter is at the expense of a bias, but which can be controlled by the kernel width that is chosen to balance over estimation bias and variance. See \eqref{eq:scott} a simple choice used in this work. 
The theory and Monte Carlo method have been investigated in a number of examples, illustrating sensitivity parameters of different nature that prompt attention, e.g., $\hp_3$ in Example~1 (Normal response) has non-zero response gradient but zero sensitivity; $\hp_2$ in Example~2 (buckling) has a significant chance of zero response gradient; $\hp_2$ in Example~3 (first passage) displays weak correlation between response and its gradient but still carries significant reliability sensitivity; and $\hp_1$ (pile diameter) whose sensitivity is intuitive but not trivial.

%-----------------------------------------------------
\subsection{Limitations and future work}
%-----------------------------------------------------
Limitations and potential future work are in order. 
%----------------------
% smoothness
%----------------------
The theory assumes the response function to be sufficiently smooth to justify the existence of gradient and continuity of their joint/marginal PDFs.
%-----------------------------
% variance reduction
%-----------------------------
The Monte Carlo method has not exploited the relationship between the response and its gradient for variance reduction, which can be a future topic. 
%--------------------------------------------
% other methods
%--------------------------------------------
The use of kernel smoothing was motivated by the integral form in~\eqref{eq:dF_intg}, but it is not the only means. 
For example, the expectation form in \eqref{eq:dF_E} motivates the general perspective of regressing $\partial Y/\partial \hp$ against $Y$, which opens up the possibility of data analytics such as Gaussian process regression. 
In this regard, the input random variables $\bX$ may be seen as hidden variables, and it may be possible to incorporate their information for better estimating the conditional expectation in \eqref{eq:dF_E}. 
%----------------------------
% 2nd derivatives
%----------------------------
On extension of theory, it should be possible to obtain the second derivative $\partial^2 P(Y\ge y)/\partial \hp_i \partial \hp_j$. The derivation should be similar to Section~7.3 of \cite{Au2026_dR1} and the result should be similar to (if not the same as) (13) there. 
It has not been done in this work because there are questions that deserve a separate study, e.g., how/whether second derivatives are used in applications, and the required precision; whether kernel smoothing can still be used to resolve zero-probability conditioning, especially for the derivative term in (13) there.

%-----------------------------------------------------
\subsection{Future vision with gradient}
%-----------------------------------------------------
This work is part of an effort to explore the use of gradient information for reliability calculations and related objectives, hoping to break bottlenecks when only response values are used. 
Such information may be limited in the past, but can become a norm in the near future, thanks to the advent of computer hardware (e.g., GPU), computer codes (e.g., Pytorch) and increasingly frequent use of neural network models.
A future vision is that system response will be computed together with gradients in an embedeed manner for various purposes, e.g., reliability sensitivity, model updating, training surrogate models, and design optimization. This calls for a vitalization in reliability methodologies to be in par with future response calculation codes. 
%

%===============================================
\section{Acknowledgments}
%===============================================
The research presented in this paper is supported by Academic Research Fund Tier 1 (RG68/22) from the Ministry of Education, Singapore. Any opinions, findings and conclusions or recommendations expressed in this material are those of the authors and do not reflect the views of the funders.

\appendix

%{\color{red}
%%================================================
%\section{Analytical solution of $P(Y<y)$ for shear building buckling}\label{sec:buckle_proof}
%%================================================
%}

%================================================
% to generate .bbl file, use \bibliography{export} then run latex, then bibtex
%================================================
\bibliographystyle{elsarticle-num}

%\bibliography{sensitivity.bbl}
%\bibliography{export}

%==============================================

%==============================================

\end{document}